\documentclass[%
 aip,
rsi,%
 amsmath,amssymb,
 reprint,%
]{revtex4-1}

\usepackage{graphicx,epsfig}
\usepackage{dcolumn}
\usepackage{theorem,epsf,psfrag,verbatim}
\usepackage{array,float}
\usepackage{color}
\usepackage{amsmath,amssymb,bm}

\newcommand{\obs}[1]{{\color{black}{#1}}}

\def\fbu{\hat {\bm u}}

\def\hm{{\bm h}^-_{\bm k}}
\def\hp{{\bm h}^+_{\bm k}}
\def\hpm{{\bm h}^\pm_{\bm k}}
\def\hsk{{\bm h}^{s_k}_{\bm k}}
\def\hsp{{\bm h}^{s_p}_{\bm p}}
\def\hsq{{\bm h}^{s_q}_{\bm q}}
\def\bu{{\bm u}}
\def\hperpk{{\bm h}^{2D}_{\bm k}}
\def\hpark{{\bm h}^{\theta}_{\bm k}}
\def\hparsk{{\bm h}^{s_k,\theta}_{\bm k}}
\def\hparpk{{\bm h}^{+,\theta}_{\bm k}}
\def\hparmk{{\bm h}^{-,\theta}_{\bm k}}

\def\p{\partial}
\def\eps{\varepsilon}

\newcommand{\be}{\begin{equation}}

\newcommand{\ee}{\end{equation}}
\newcommand{\bk}{{\bm k}}
\newcommand{\bx}{{\bm x}}

\newcommand{\bz}{{\bm z}}
\newcommand{\bv}{{\bm v}}
\newcommand{\bF}{{\bm f}}
\newcommand{\bp}{{\bm p}}
\newcommand{\bq}{{\bm q}}
\newcommand{\buperp}{{\bm u}^{\rm 2D}}
\newcommand{\fbuperp}{\hat{\bm u}^{\rm 2D}}
\newcommand{\bomega}{{\bm \omega}}
\newcommand{\btheta}{{\bm \theta}}
\newcommand{\bomperp}{{\bm \omega}^{\theta}}
\newcommand{\bompar}{\omega \hat{\bm z}}
\newcommand{\ompar}{{\omega}}

\newcommand{\fpsik}{\hat \psi_{\bm k}}
\newcommand{\fpsip}{\hat \psi_{\bm p}}

\newcommand{\fpsiq}{\hat \psi_{\bm q}}
\newcommand{\fthetak}{\hat \theta_{\bm k}}
\newcommand{\fthetap}{\hat \theta_{\bm p}}
\newcommand{\fthetaq}{\hat \theta_{\bm q}}

\begin{document}
\vspace{-3em}
\begin{minipage}[l]{\textwidth}
\noindent
Postprint version of the manuscript published in Physics of Fluids {\bf 29}, 111101 (2017). \\
\end{minipage}
\vspace{1em}

\title[2D3C flows]{From two-dimensional to three-dimensional turbulence through two-dimensional three-component flows}

\author{L. Biferale}
 \email{biferale@roma2.infn.it}
\author{M. Buzzicotti}
\author{M. Linkmann}%
 \affiliation{Department of Physics \& INFN, University of Rome `Tor Vergata', Via della Ricerca Scientifica 1, 00133 Rome, Italy}
\date{\today}

\begin{abstract}
  The relevance of two-dimensional three-components (2D3C) flows goes well beyond their  occurrence in nature, and 
a deeper understanding of their dynamics might be  also helpful 
in order to shed further light on the dynamics of pure two-dimenional (2D) or three-dimensional (3D) flows and vice versa.
The purpose of the present paper is to make a step in this direction through 
a combination of numerical and analytical work. The analytical part is mainly 
concerned with the behavior of 2D3C flows in isolation and the connection between the geometry 
of the nonlinear interactions and the resulting energy transfer directions. 
Special emphasis is given to the role of helicity. We show that a generic 2D3C flow can be described by two stream 
functions corresponding to the two helical sectors of the velocity field. 
The projection onto one helical sector (homochiral flow) leads to a full 3D constraint and to the inviscid conservation 
of the total (three dimensional) enstrophy and hence to an inverse cascade of the kinetic energy of the third component also.  
The coupling between several 2D3C flows is studied through a set of suitably
designed direct numerical simulations (DNS), where we also explore the transition
between 2D and fully 3D turbulence.
{
In particular, we find that the coupling of three 2D3C
flows on mutually orthogonal planes subject to small-scale forcing
leads to stationary 3D out-of-equilibrium dynamics at the energy containing scales.
}
The transition between 2D and 3D turbulence is then explored through adding a 
percentage of fully 3D Fourier modes in the volume.
  %
\end{abstract}

\maketitle

\section{Introduction}
In his 1967 paper,  Kraichnan\cite{Kraichnan67} provided the basis of the current understanding of  
2D turbulence. His argument explained  
the existence of two mutually exclusive inertial ranges  
corresponding to an inverse energy cascade with an $E(k)\sim k^{-5/3}$ energy spectrum,
 and to a direct enstrophy cascade with $E(k) \sim k^{-3}$, as well 
as the formation of a large-scale condensate. The dynamics of 
the two cascades were predicted to have some important differences, whereby the energy should cascade
locally in Fourier space while the transfer of enstrophy proceeds mainly through nonlocal interactions.   
Both cascades have been  observed in experiments and numerical simulations,
see e.g. Refs.~\onlinecite{Lilly69,Sommeria86,Smith93,Paret97,Gotoh98,Chen03,Lindborg00}, 
and so has the formation of a large-scale condensate in the form of two 
counter-rotating vortices \cite{Hossain83,Smith93,Smith94,Paret98,Boffetta10}
(see also the recent review in Ref.~\onlinecite{Boffetta12} and references therein).
On the other hand, pure 3D flows have a completely
different phenomenology. They develop a forward energy cascade and a build-up of
non-Gaussian fluctuations at successively smaller scales \cite{Frisch95}. Remarkably
enough, there are many instances in nature when the flow is neither 2D nor 3D,
enjoying a forward or an inverse energy cascade (or both) depending on the boundary
conditions as for the case of thick turbulent layers
\cite{Celani10,Xia11,Nastrom84,Benavides17}, or on the existence of a strong
rotation rate \obs{\cite{Cambon89,Waleffe93,Smith99,Chen05,Mininni09,Gallet15a,BiferalePRX2016}}, or on the presence of a strong
mean magnetic field \cite{Moffatt67,Alemany79,Zikanov98,Gallet15,Alexakis11,Bigot11} for 
conducting flows. In all the above instances, the inverse cascade is
triggered by a tendency of the flow to become a two-dimensional three-components (2D3C) flow  
where the third component is passively advected by the
two-dimensional flow and the physics 
is dominated by the inverse energy cascade in the 2D plane. 
Interestingly enough, recent numerical simulations have also shown
that an inverse energy cascade can be sustained by a fully 3D isotropic flow if
constrained to evolve only on the subset of 
homochiral helical Fourier waves
\cite{Waleffe92,Biferale12,Alexakis17}, 
\obs{i.e., those of like-signed helicity}. 
In other words there might exist fully 3D
structures that bring energy upscale,  contrary to what is 
typically observed for unconstrained  3D  turbulence. 
\\
\noindent
Apart from the above-mentioned applications,  2D3C 
flows might also be relevant for the evolution of fully three-dimensional 
homogeneous turbulence \cite{Waleffe92,Moffatt14}. 
The connection lies in the structure of the inertial term $(\bu \cdot \nabla) \bu$ of the 
Navier-Stokes equations that in Fourier space can be written as the superposition of interactions among three wavevectors $\bk, \bp$ and $\bq$
which satisfy the condition to  form a closed triad: $\bk +\bp+\bq=0$.
\obs{Any} triad defines a plane in Fourier 
space, which \obs{through a suitable choice of coordinate system} 
can be \obs{identified with}, e.g., the \obs{$(k_x,k_y)$}-plane.
The   
flow corresponding to this single triad of wavevectors is therefore 
independent of the $z$-coordinate \cite{Moffatt14}.  Of course, the superposition of many unoriented  2D3C triads is not
a 2D3C flows. Hence, there is no {\em a priori} reason 
to believe that the full turbulent evolution of a quasi-isotropic
3D flows has anything in common with the limiting case of a set of co-planar 2D3C triads \cite{Moffatt14}. \\
We can  summarize by saying that we know at least two different mechanisms which might trigger a reversal
of the energy cascade in a 3D turbulent flow:  either the system is pushed toward a 2D (or 2D3C) configuration
by some anisotropic external mechanism or it must be constrained to evolve  on a strongly helical manifold,
with breaking of mirror (but not rotational) symmetry at all scales. The two mechanisms are somehow opposite,
helicity being a fully three dimensional quantity. \\
\obs{In this paper we investigate through analytical and numerical methods}
how one might smoothly go from a pure 2D3C dynamics to a 3D dynamics
by successive addition of triads under some
controlled protocol. The main interest is to understand how/when the system starts 
to show 3D behavior and 
what the main physical mechanisms are that trigger the transition. 
{ 
For this purpose we first analyze the 2D3C dynamics, which in fact is
characterized by a split energy cascade, where the passively advected third
component undergoes a direct energy cascade while the advecting flow
shows an inverse energy cascade.
In order to better understand the split cascade we express the velocity field in two 
different bases. We use either  
}
the standard decomposition of a pure 2D flow plus
a component passively advected or the 
decomposition in positive and negative helical Fourier waves \cite{Constantin88,Waleffe92}
that will be the building block of any 3D dynamics. 
{
Understanding the relation between the two bases will prove helpful in the
description of the physics of the transition from 2D3C to 3D behavior, 
as the characteristic split cascade of a 2D3C flow will be 
obscured in the presence of 3D dynamics.}
\\
In order to understand the transition to a 3D flow we first
study the dynamics of a flow that evolves on three 2D3C manifolds defined
on mutually orthogonal planes. 
As a result we obtain a flow that has the discrete rotational symmetries of the cube
and which is mainly based on three weakly coupled 2D3C evolutions. 
Successively we start to add a percentage $0 \leqslant \alpha \leqslant 1$ of Fourier modes
randomly (but quenched in time) in the whole 3D cube and we study the transition
to a full 3D isotropic dynamics at increasing $\alpha \rightarrow 1$. 
We show that the basic 2D3C flow can be described by introducing
two independent stream functions, $\psi^+,\psi^-$, which are  connected to its helical decomposition. We also found that imposing the 
homochiral constraint on the dynamics
will  force one of the two stream functions to be identically zero and the
2D3C dynamics to collapse on a constrained configuration where the 
component out of the plane is not a 
simple passive scalar anymore, 
and that this has important consequences for the global energy transfer. 
Finally, we study the transition from 2D to 3D dynamics
as a function of the percentage $\alpha$ of added 3D modes,
and we  assess the dynamical relevance of the energy transfer 
due to homochiral triads in the fully 2D3C configuration and 
for different $\alpha$.
\\

\noindent
This paper is organized as follows. In section \ref{sec:basics} we discuss 
the basic setup and review the inviscid invariants particular to 2D3C flows as
derived in Ref.~\onlinecite{Moffatt14}. Section \ref{sec:helical_dec} contains the main 
part of the theoretical work on the helical decomposition of 2D3C flows, where 
the description of a 2D3C flow in terms of two stream functions is introduced and 
the helical decomposition is used to study the effect of different helical interactions
on the dynamics of the planar and perpendicular components of the velocity field. 
The numerical simulations are described in Section \ref{sec:simulations}, beginning with a comparison between 
the dynamics of  single and coupled 2D3C flows. 
Subsequently we investigate the transition from 2D to 3D turbulence in Section \ref{sec:2d3d-transition} and 
the behavior of a subset of helical interactions leading to an inverse energy 
cascade in Section \ref{sec:homochiral_numerics}. 
We summarize our results in Section \ref{sec:conclusions}. 
\\  

\section{Structure and inviscid invariants of the 2D3C Navier-Stokes equations}
\label{sec:basics}
\noindent
We start from considering  a 2D3C solenoidal velocity field $\bu=(u_x,u_y,u_z)$ on a three-dimensional
domain $V=[0,L)^3$ with periodic boundary conditions, where $u_x$, $u_y$ and $u_z$
  are functions of only $x$ and $y$.
We define
the 2D-component $\buperp$ and the component perpendicular to the $(x,y)$-plane
as $\btheta$ to stress that it behaves as a passive scalar (see below)
\be
\label{eq:2d3c-decomp-u}
\buperp = \begin{pmatrix}u_x \\u_y \\ 0   \end{pmatrix}
\qquad \mbox{and} \qquad
\btheta =  \begin{pmatrix} 0  \\ 0  \\ u_z \end{pmatrix} \ .
\ee
such that the total vector field is given by $\bu = \buperp + \btheta$.
For such a 2D3C flow, the 3D Navier-Stokes equations split into the 2D
Navier-Stokes equations for $\buperp$, while $\theta \equiv u_z$ is passively advected
\begin{align}
\label{eq:2d3c-nse}
\p_t \buperp &= -(\buperp \cdot \nabla) \buperp - \nabla P + \nu \Delta \buperp \ ,\nonumber \\  
\p_t \theta &= -(\buperp \cdot \nabla) \theta + \nu \Delta \theta  \ , 
\end{align}
where $P$ is the two dimensional pressure and $\nu$ the kinematic viscosity.
For the vorticity $\bomega =  \nabla \times \bu$ the decomposition of
$\bu$ into $\buperp$ and $\btheta$ results in
\begin{align}
\label{eq:2d3c-decomp-omega}
\bomperp = \nabla \times \btheta
         &= \begin{pmatrix} \p_y \theta \\ -\p_x \theta \\ 0   \end{pmatrix}
         = \begin{pmatrix}\omega_x \\ \omega_y \\ 0   \end{pmatrix} \ , \\
        \nabla \times \buperp
        &= \begin{pmatrix} 0  \\ 0  \\ \p_x u_y - \p_y u_x \end{pmatrix}
        = \begin{pmatrix} 0  \\ 0  \\ \omega_z \end{pmatrix} \ ,
\end{align} 
and for simplicity we define $\ompar \equiv \omega_z$ 
such that the total vorticity vector field is $\bomega = \bomperp + \bompar$, 
with $\hat{\bm z}$ denoting the unit vector in the $z$-direction. 
The 2D3C Navier-Stokes equations \eqref{eq:2d3c-nse} in the
vorticity formulation then become
\begin{align}
\label{eq:2d3c-nse-vort}
\p_t \bomperp &= -(\buperp \cdot \nabla) \bomperp + (\bomperp \cdot \nabla) \buperp + \nu \Delta \bomperp \ ,\nonumber \\  
\p_t \ompar &= -(\buperp \cdot \nabla) \ompar + \nu \Delta \ompar  \ . 
\end{align}

\noindent
The 3D Navier-Stokes equations have two inviscid quadratic invariants, the total
energy 
\be
E=\frac{\langle\bu \cdot \bu\rangle}{2} = \frac{1}{2|V|}\int_V d\bx \ |\bu|^2 \ ,
\ee
and the total kinetic helicity
\be
H=\langle\bu\cdot\bomega\rangle = \frac{1}{|V|}\int_V d\bx \ \bu\cdot \bomega \ ,
\ee 
per unit volume, while the 2D equations have two quadratic invariants, the total 2D energy
\be
E^{2D} = \frac{\langle|\buperp|^2\rangle}{2} = \frac{1}{2|V|}\int_V d\bx \ |\buperp|^2 \ , 
\ee
and the 2D enstrophy:
\be
\Omega  =  \langle|\ompar|^2\rangle = \frac{1}{|V|}\int_V d\bx \ |\ompar|^2 \ .
\ee
The 2D3C flow being 
a particular 3D case, it is immediately clear from Eqs.~\eqref{eq:2d3c-nse} and \eqref{eq:2d3c-nse-vort}  that  the energy of the passive component is  another
quadratic invariant for this case 
\be
E^\theta = 
\frac{\langle|\theta|^2\rangle}{2} = \frac{1}{2|V|}\int_V d\bx \ |\theta|^2 \ . 
\ee
When discussing the energy transfer of the 2D3C flow it is necessary to distinguish the component of the energy in the plane, $E^{2D}$, from the component out of the plane
$E^\theta$. The former is known to develop an inverse cascade, while the latter is typically transferred to small scales. Hence,
the transfer of the total energy can be either forward or backward.
For later purpose it is important to remark here that the above conservations hold in Fourier space also
on a  triad-by-triad basis \cite{Moffatt14}. 
Helicity for the 2D3C case must play a passive role
concerning the independent 2D dynamics. 
Indeed, it is easy to realize that 
it can be further decomposed in two quantities
\be
\bu \cdot \bomega = (\buperp + \btheta) \cdot (\bomperp + \bompar)
= \buperp \cdot \bomperp + \theta\ompar \ ,
\ee
since $\btheta \cdot \bomperp =0$ and $\buperp \cdot \bompar =0$.
Using Eqs.~\eqref{eq:2d3c-nse} and \eqref{eq:2d3c-nse-vort} it can also be shown
that $H_{x,y}\equiv\langle \buperp \cdot \bomperp \rangle$ and $H_{z}\equiv \langle \theta\ompar\rangle$
 are conserved and related to each other by a geometrical constraint
\begin{align}
H_{x,y} &= \langle u_x \p_y\theta -u_y \p_x\theta  \rangle 
          = -\langle \theta (\p_yu_x -\p_xu_y)  \rangle \nonumber \\
          &= \langle \theta\ompar\rangle = H_{z} \ ,
\end{align}
since $\langle \p_y(u_x \theta) - \p_x(u_y \theta) \rangle=0$ due to the periodic boundary conditions.
%
Hence, a 2D3C-flow has four inviscid \obs{quadratic} invariants, where for one of which, the total helicity,
the planar and perpendicular components are identical 
\cite{Moffatt14} and passive concerning the evolution of $\buperp$. 
%
 
\section{Helical decomposition of a 2D3C flow} \label{sec:helical_dec}
Even though helicity is a passive quantity for a pure 
2D3C dynamics, it is useful to 
further disentangle its dynamics in view of the possibility to
build up a full 3D flow by adding different 2D3C submanifolds. 
To do that, we exploit the decomposition of any incompressible 3D flow 
into helical modes by the procedure proposed in Refs. \onlinecite{Constantin88,Waleffe92}. 
Since $\bu$ is a solenoidal vector field, its Fourier modes $\fbu$ have 
only  two degrees of freedom, and we have 
\be
\fbu_\bk(t) = \fbu_\bk^+(t) + \fbu_\bk^-(t) = \hat u^+_\bk(t)\hp + \hat u^-_\bk(t)\hm \ ,
\ee
where $\hpm$ are normalized eigenvectors of the curl operator in Fourier space
\cite{Constantin88,Waleffe92}.
The helical decomposition thus decomposes the Fourier modes of the 
velocity field into two components, each of which satisfies    
\be
i\bk \times \fbu_\bk^{s_k}= s_kk \fbu_\bk^{s_k} \ ,
\ee
and $s_k = \pm$.
For a 2D3C flow this requirement  becomes
\begin{align}
i \begin{pmatrix} k_y \hat u_z^{s_k} \\ -k_x \hat u_z^{s_k} \\ k_x \hat u_y^{s_k} - k_y \hat u_x^{s_k} \end{pmatrix}  
 = 
s_kk  \begin{pmatrix} \hat u_x^{s_k} \\ \hat u_y^{s_k} \\ \hat u_z^{s_k} \end{pmatrix}
\ ,
\end{align}
and we obtain
\be
\label{eq:helical_dec_2d3c}
\fbu_\bk^{s_k} = s_k \hat u_z^{s_k} \begin{pmatrix} i{k_y}/{k}  \\  -i{k_x}/{k} \\ \obs{s_k} \end{pmatrix} \ ,
\ee
\obs{where the symbol $s_k$ in the third component of $\fbu_\bk^{s_k}$ 
stands for $\pm1$.}
\subsection{Introduction of two stream functions}
For each helical sector we can now define two stream functions
through their respective Fourier transforms
\be
 \fpsik^+ \equiv \hat u_z^{+}/k \qquad \mbox{and} \qquad \fpsik^- \equiv -\hat u_z^{-}/k \ ,
\ee
such that $\fpsik^{s_k}$ are Hermitian-symmetric and
\be
\label{eq:helical_dec_psi}
\fbu_\bk^{s_k} = k\fpsik^{s_k} \hsk \ ,
\ee
where the helical basis vectors are given as
\label{eq:basis_2d3c}
\be
\hsk = \begin{pmatrix} i{k_y}/{k}  \\  -i{k_x}/{k} \\ \obs{s_k} \end{pmatrix} \ .
\ee
A 2D3C-flow can therefore be described in real space by two stream functions $\psi^+$ and $\psi^-$
\be
\bu= \begin{pmatrix}\p_y (\psi^+ + \psi^-) \\ -\p_x (\psi^+ + \psi^-)  \\ (-\Delta)^{1/2}(\psi^+ - \psi^-) \end{pmatrix} \ ,
\ee
such that
\begin{align}
\buperp&= \begin{pmatrix}\p_y (\psi^+ + \psi^-) \\ -\p_x (\psi^+ + \psi^-)  \\ 0 \end{pmatrix} \ ,
\label{eq:uperp_psi_pm}
\\
\theta &=  (-\Delta)^{1/2}(\psi^+ - \psi^-) \ .
\label{eq:upar_psi_pm}
\end{align}
It is important to stress that we are just changing basis, we move from the usual description in terms of the couple of functions $(\psi,\theta)$, describing the stream function
of the $\buperp$  field and the out-of-plane passive component to a couple of fully 3D structures $(\psi^+,\psi^-)$  that also reconstruct the original 2D3C flow.
%
While the formulation $(\psi,\theta)$ is
natural for the study of turbulence under rapid rotation, passive scalar
advection in 2D turbulence or for thick layers of fluid, the helical
formulation $(\psi^+,\psi^-)$ is the natural decomposition for fully 3D flows.
 \noindent
Clearly, there exist two possibilities to 
reduce the complexity of a 2D3C flow, either through 
requiring (i) $\psi^+=\psi^-$, or, (ii) by setting one of the helical stream function to zero, $\psi^-=0$, say.
The entire evolution of the flow in both cases is given by one stream function only, and the quantities 
$\buperp$, $\theta$, $\bomperp$ and $\omega$ are obtained 
through taking the appropriate derivatives 
of the single stream function.   
In the first case where $\psi^+=\psi^-$ the flow is fully 2D as $\theta = 0$, 
and the inner product $\bu \cdot \bomega$ then vanishes identically.
That is, a 2D3C flow with vanishing pointwise helicity is a 2D flow.
In the second case where $\psi^-=0$, say, the  
inner product $\bu \cdot \bomega = \omega \theta \neq 0$ 
at all times, hence the flow must be 3D and $\theta$ correlates with $\omega$ at all times.
The two cases bear the important difference that the latter case is 
dynamically unstable while the 
former is not. The flow in case (i) which is initially 2D will remain so unless it is subjected to 
3D perturbations, while in case (ii) negatively helical Fourier modes are 
always created through nonlinear interactions.    
In order to maintain a flow in case (ii) where $\psi^- =0$ at all times, it is necessary to 
re-project the evolution onto the submanifold corresponding to $\psi^+$ alone.  
\\

\noindent
Separate evolution equations for $\psi^+$ and $\psi^-$ can be 
obtained from the respective evolution equations for $\theta = (-\Delta)^{1/2}(\psi^+ - \psi^-)$ 
and $\ompar = -\Delta(\psi^+ + \psi^-)$ 
(we neglect the dissipative terms from now on)
\begin{align}
\label{eq:evol_psi_pm}
\p_t &\psi^\pm =   
\mp\frac{(-\Delta)^{-1/2}}{2}
 \hspace{-1em}\sum_{s \in \{+,-\}}\hspace{-1em} 
s\Big[\nabla (-\Delta)^{1/2}\psi^s \times \nabla( \psi^s +\psi^{-s}) \Big] \nonumber \\  
& -\frac{(-\Delta)^{-1}}{2} 
 \hspace{-1em}\sum_{s \in \{+,-\}}\hspace{-1em} 
\Big[
\nabla (-\Delta)\psi^s \times \nabla (\psi^s + \psi^{-s}) \Big] ; 
\end{align}
which leads to the usual  equation
of the 2D stream function $\psi^{2D}$  when $\psi^+ = \psi^-$   
\be
\label{eq:evol_psi2D}
\p_t \psi^{2D} = 
(-\Delta)^{-1}(\nabla \psi^{2D} \times \nabla (-\Delta) \psi^{2D})_z \ .
\ee
Also in the fully helical case,  $\psi^-=0$, the flow is again described by one stream function only. 
However, the evolution of the single stream function 
$\psi^+$ derived from Eq.~\eqref{eq:evol_psi_pm} 
differs in this case from that of a 2D flow given by Eq.~\eqref{eq:evol_psi2D}
\begin{align}
\label{eq:psiplus}
\p_t \psi^+ = &  
 \ \frac{(-\Delta)^{-1/2}}{2}(\nabla \psi^+ \times \nabla(-\Delta)^{1/2} \psi^+ )_z \nonumber \\
& +\frac{(-\Delta)^{-1}}{2}(\nabla \psi^+ \times \nabla(- \Delta) \psi^+)_z \ .
\end{align}
As discussed before, the removal of one degree of freedom in the helical decomposition 
forces the perpendicular component $\theta$ to be correlated to the 2D vorticity $\omega$. 
In this case  $\theta$ does not evolve anymore as a passive scalar 
%
with  important
consequences for the cascade direction of its energy $E_\theta$, as discussed in the following section. 
  
\subsection{Cascade directions and helical interactions}
\label{sec:cascade_directions}
The inviscid invariants can be expressed in terms of the Fourier transforms of $\psi^+$ and $\psi^-$
\begin{align}
\label{eq:inv-u2d}E^{2D} &= \frac{1}{2}\sum_{\bk \in \mathbb{Z}^3}  k^2|\fpsik^+ +\fpsik^-|^2 \ , \\ 
\label{eq:inv-theta}
E^\theta &= \frac{1}{2}\sum_{\bk \in \mathbb{Z}^3}  k^2|\fpsik^+ - \fpsik^-|^2 \ , \\ 
\label{eq:inv-omega2d}
\Omega &= \sum_{\bk \in \mathbb{Z}^3}  k^4|\fpsik^+ + \fpsik^-|^2 \ , \\ 
H_z &= \sum_{\bk \in \mathbb{Z}^3}  k^3(|\fpsik^+|^2 - |\fpsik^-|^2) \ , 
\end{align}
while the enstrophy in the plane can be written as
\be
\langle|\bomperp|^2\rangle = \sum_{\bk \in \mathbb{Z}^3}  k^4|\fpsik^+ -\fpsik^-|^2 \ .
\ee
For a fully helical velocity field ($\fpsik^-=0$), we have that $E^{2D} = E^\theta$ and
$ \langle|\bomperp|^2\rangle = \Omega $. As a result also the `enstrophy' of the $\theta$ field is conserved leading to
an inversion of the direction of the $E^\theta$ transfer. 
Thus, for 2D3C homochiral evolutions, 
the total energy 
$E=E^{2D} + E^\theta$
will necessarily be transferred upscale because also 
the `passive' component does not behave anymore as a passive scalar in 2D  and will develop
an inverse energy cascade. 
%
The latter observation also implies that 
a forward cascade of $E^\theta$ in the full 2D3C case 
can only occur through interactions between $\psi^+$ and $\psi^-$ (see also Appendix \ref{app:passive_scalar}). 
The above 
energy transfer 
property of homochiral triads was
already discussed in the original paper by Waleffe \cite{Waleffe92} and it is at the basis of the numerical simulations of homochiral turbulence
developed in Refs.~\onlinecite{Biferale12,Biferale13,Sahoo15} 
resulting in 
3D fully isotropic turbulence with an inverse cascade. Here we stress the connection with the
underlying 2D3C structure of any isolated triadic Navier-Stokes  interaction and the important remark that in such a case the direction of the transfer of the total energy can always be decomposed in two contributions, one due to the transfer of the 2D physics (always inverse) and one due to the amount of energy transferred by the out-of-plane components, which is typically forward
and only backward if we restrict to homochiral dynamics. 

\obs{
In this context it is interesting to notice 
that in a rotating fluid Ekman pumping near a 
solid boundary leads to a vertical velocity component which is directly 
proportional to the vertical vorticity component, that is, to a flow with 
pointwise positive helicity everywhere close to the boundary. 
The result is an effective reduction in the number of degrees of 
freedom, because in terms of the helical decomposition the 
resulting flow would be described
by one helical stream function 
only. In consequence, the vertical velocity component 
close to the boundary in a rapidly rotating flow should display an 
inverse energy cascade. 
}

\subsection{Fourier decomposition}
\label{sec:Fourier}
In order to shed some further light on the couplings between the two stream functions, 
their respective contributions to the 2D dynamics and the dynamics of the 
perpendicular component, 
we consider the evolution  of the Fourier transforms of 
$\psi^+$ and $\psi^-$
\begin{align}
\label{eq:Fevolution_psipm}
\p_t &(k \sum_{s_k}\fpsik^{s_k}\hsk)^* =-(i\bk \hat{ \mathcal{P}})^* \nonumber \\ 
&-\frac{1}{2}
\sum_{\bk+ \bp+\bq =0} \sum_{s_p,s_q}\fpsip^{s_p}\fpsiq^{s_q}pq(s_pp-s_qq) \left(\hsp \times \hsq \right) \ ,
\end{align}
which are obtained directly from the Navier-Stokes equations in Fourier space by first substituting the general
helical decomposition for a 3D velocity field \cite{Waleffe92} and subsequently using 
Eq.~\eqref{eq:helical_dec_psi}. Here $\mathcal{P} \equiv P + |\bu|^2/2$, 
as the inertial term has been written in rotational form.
\begin{figure*}
\begin{center}
\includegraphics[width=0.5\columnwidth]{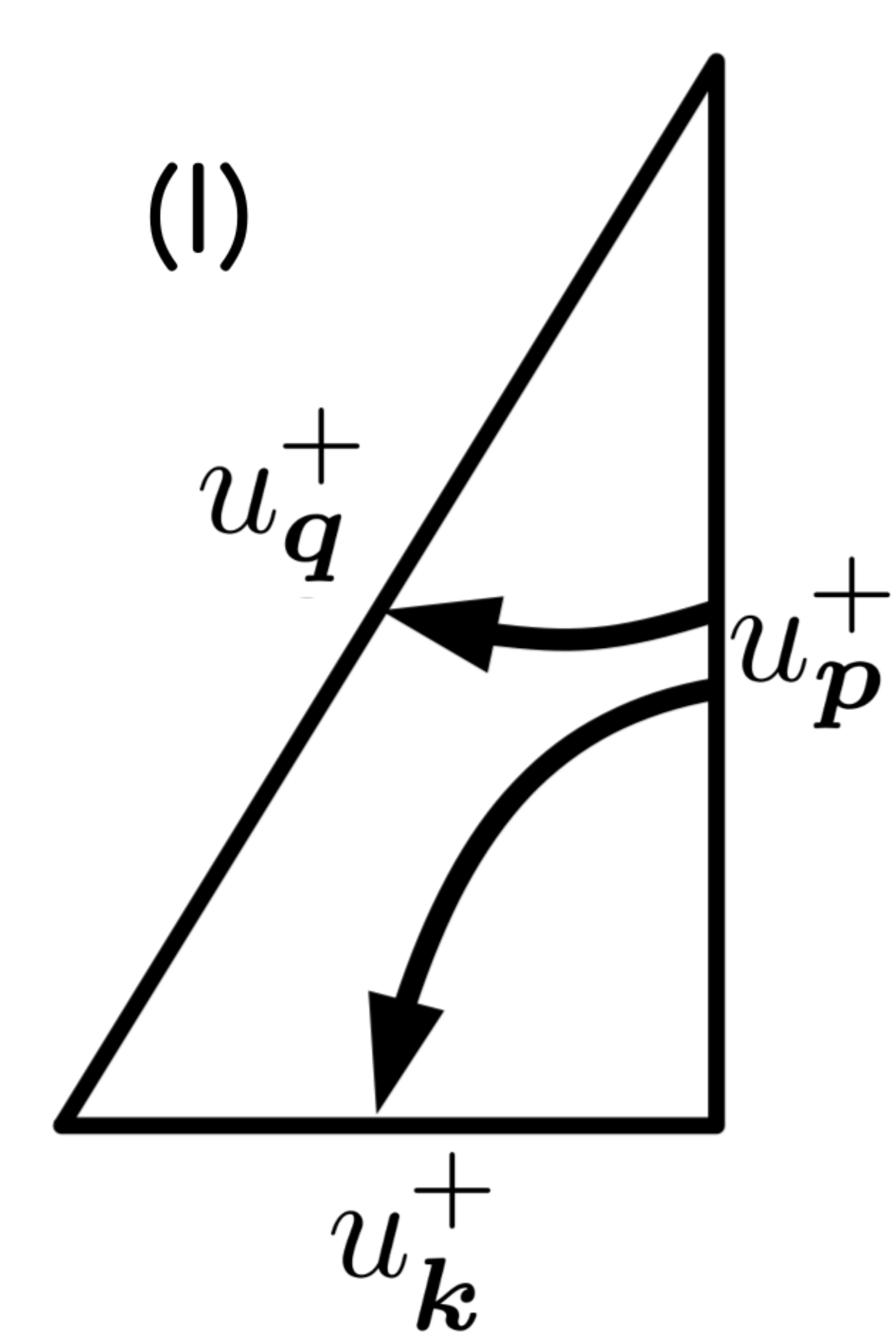}
\includegraphics[width=0.5\columnwidth]{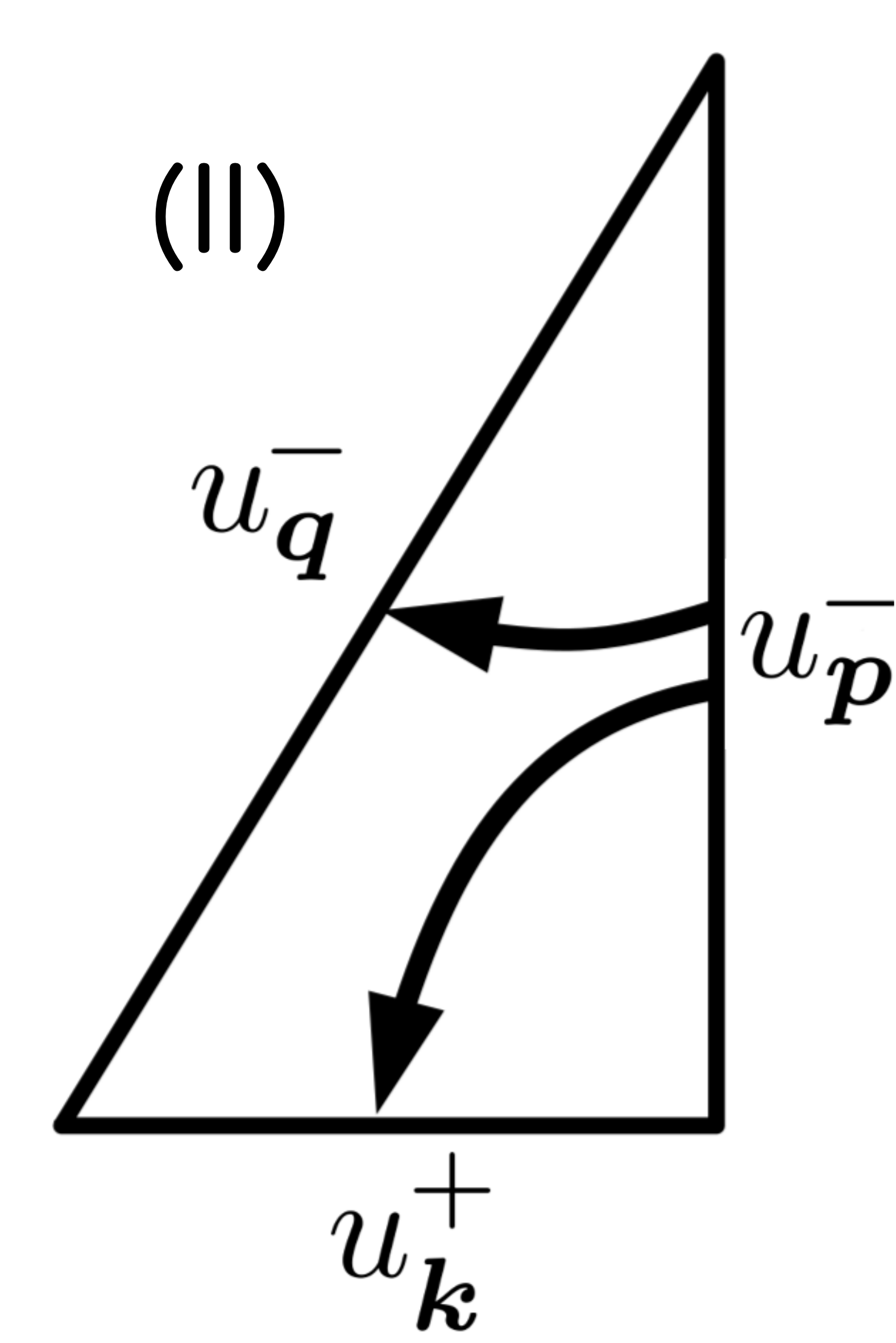}
\includegraphics[width=0.5\columnwidth]{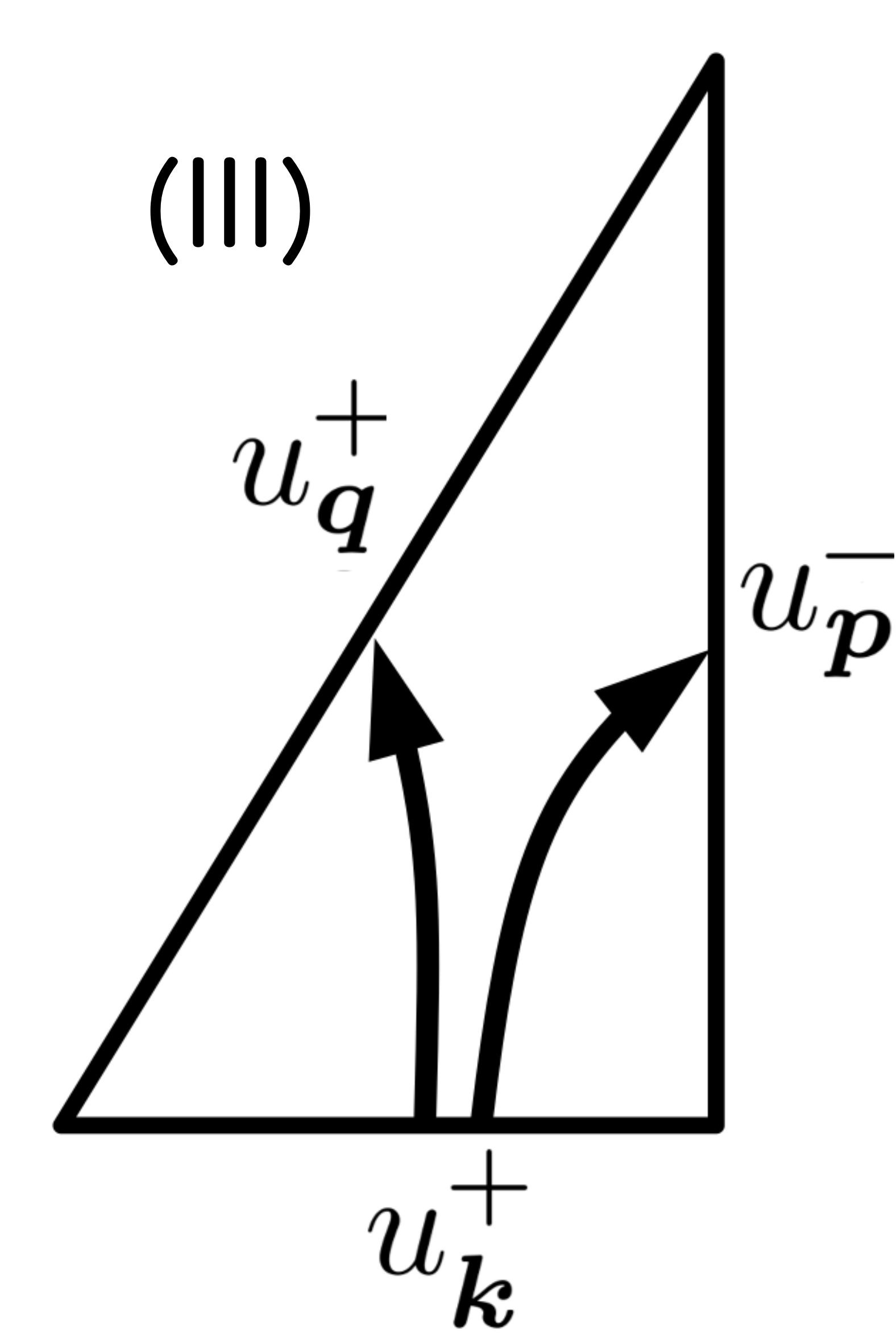}
\includegraphics[width=0.5\columnwidth]{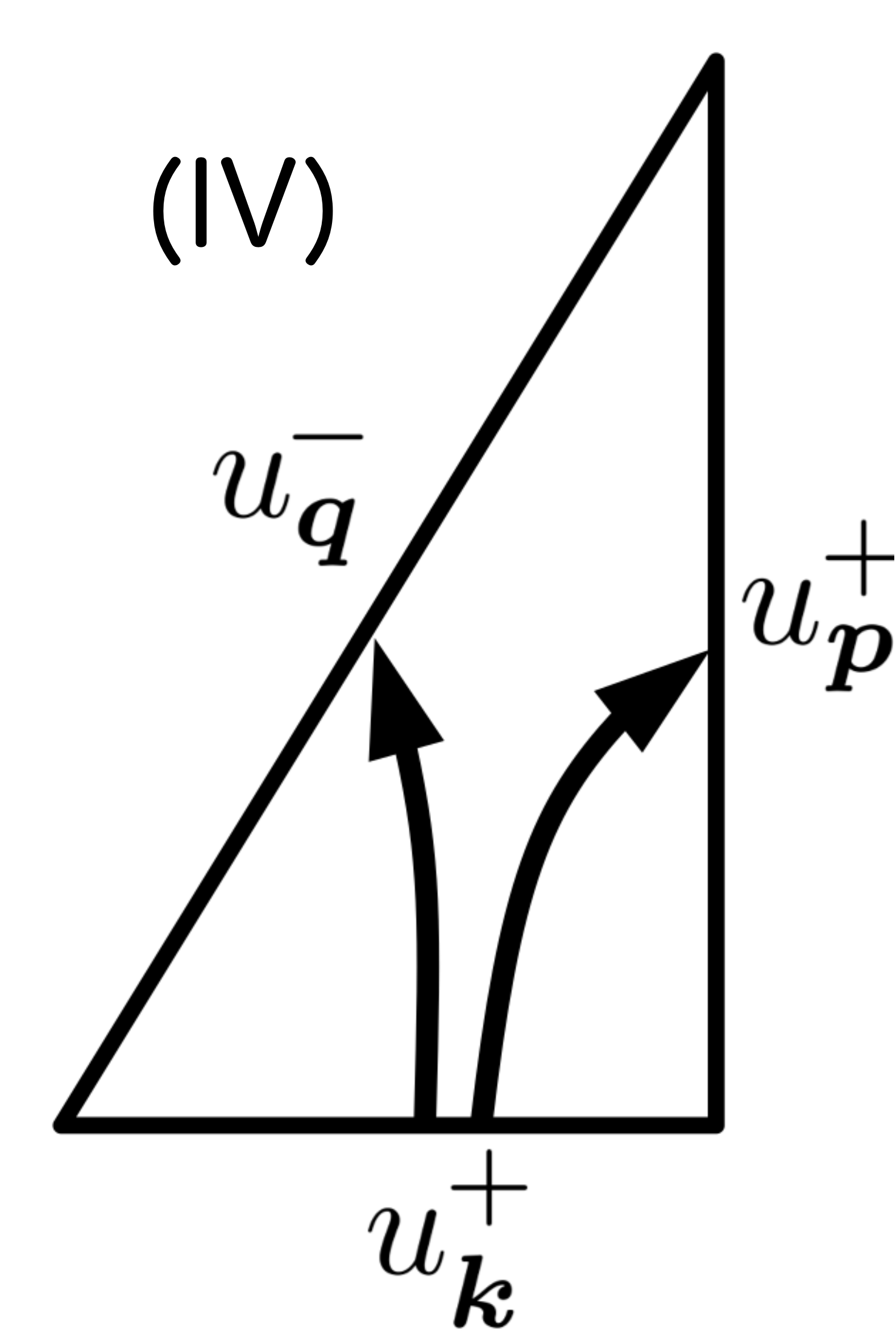}\\
\hspace{-0.5cm}$\underbrace{\hspace{3cm}}_\text{\Large homochiral}
\hspace{1.5cm}
\underbrace{\hspace{12cm}}_\text{\Large heterochiral} $ 
\caption{Classes of triad interactions according to the helical decomposition. The arrows indicate
energy transfers deduced from the stability arguments in Ref.~\onlinecite{Waleffe92}.}
\label{fig:triads}
\end{center}
\end{figure*}
After some algebra (see Appendix \ref{app:coupling_factors}) it is possible to obtain the evolution equations for
the stream functions in the helical decomposition with planar and perpendicular contributions written separately
\be
\label{eq:Fevolution_psi}
\p_t (k\fpsik^{s_k})^* = \frac{1}{4} \sum_{\bk+ \bp+\bq =0} \
\sum_{s_p,s_q} G_{s_ks_ps_q} \ \fpsip^{s_p}\fpsiq^{s_q} \ ,
\ee
since the coupling factor
\be
G_{s_ks_ps_q} = \Big(\underbrace{\frac{p^2-q^2}{k}}_\text{$(x,y)$-plane} + \underbrace{s_k(s_pp-s_qq)}_\text{$z$-component} \Big)
kp \sin\varphi_{k,p} \ , 
\ee
consists of a contribution coming from the 2D dynamics and one coming from the dynamics of the
perpendicular component.

We now consider homo- and heterochiral helical interactions separately. A
homochiral interaction has by definition  $s_k=s_p = s_q$, while a
heterochiral interaction must have both signs of helicity.  All possible
helical triad interactions present in the Navier-Stokes equations are depicted
in Fig.~\ref{fig:triads}, where without loss of generality we only consider
$s_k=+$. Homochiral interactions correspond to triads of Class (I) in
Fig.~\ref{fig:triads} and are known to produce an inverse energy cascade
\cite{Waleffe92,Biferale12,Biferale13}.  Heterochiral triads correspond to
Classes (II-IV) in Fig.~\ref{fig:triads}, where Classes (III) and (IV) lead to
a direct cascade while the energy transfer direction deduced
from Class (II) depends on the geometry of the triad.
{The peculiar behavior of Class (II) triads had already been inferred by
Waleffe\cite{Waleffe92}, it has since been confirmed in numerically using
shell models \cite{DePietro15}. Class (II) triads also possess further inviscid
invariants which are geometry-dependent and determine the direction of the energy transfer \cite{Rathmann16}.        
}

\subsubsection{Homochiral interactions}
For $s_k=s_p= s_q$ (Class (I)), the coupling 
factors become
\begin{align}
G_{+++} &= (p-q)\Big(\underbrace{\frac{p+q}{k}}_\text{$(x,y)$-plane} 
+ \underbrace{1}_\text{$z$-component} \Big) kp \sin\varphi_{k,p} , 
\end{align}
and since the triangle inequality $p+q \geqslant k$ implies that the term 
corresponding to 2D dynamics is larger than the term corresponding to the dynamics of the
perpendicular component, we conclude that homochiral interactions mainly contribute to the 2D dynamics.
The weighting of planar and 
perpendicular parts of the coupling factor 
implies that the extra term which appears 
in the homochiral case in the evolution equation \eqref{eq:psiplus} of the stream function 
is subdominant. {This suggests that a homochiral 2D3C flow 
has similar dynamics compared to a fully 2D flow, 
which is also reflected in the fact that 
despite being not fully 2D, it conserves the total enstrophy. }

\subsubsection{Heterochiral interactions}
For $s_p\neq s_q$ (Classes (III) and (IV)), the coupling 
factors become
\begin{align}
G_{++-} &= (p+q)\Big(\underbrace{\frac{p-q}{k}}_\text{$(x,y)$-plane} + \underbrace{1}_\text{$z$-component} \Big) kp \sin\varphi_{k,p} , \\
G_{+-+} &= (p+q)\Big(\underbrace{\frac{p-q}{k}}_\text{$(x,y)$-plane} - \underbrace{1}_\text{$z$-component} \Big) kp \sin\varphi_{k,p} , 
\end{align}
and since the triangle inequality $p+q \geqslant k$ implies $|p-q|\leqslant k$, the term 
corresponding to 2D dynamics is smaller in magnitude than 
that corresponding to the dynamics of the
perpendicular component. Hence we conclude that these heterochiral interactions mainly contribute to the dynamics
of the perpendicular component. \\

\noindent
The final case to consider is a heterochiral interaction of Class (II), where $s_k \neq s_p = s_q$. In this case, the 
coupling factor becomes
\be
G_{+--} = (p-q)\Big(\underbrace{\frac{p+q}{k}}_\text{$(x,y)$-plane} - \underbrace{1}_\text{$z$-component} \Big) kp \sin\varphi_{k,p} , 
\ee
and we see that this type of interaction behaves differently from the other
heterochiral interactions. Similar to the homochiral case, the triangle
inequality $p+q \geqslant k$ implies that the term corresponding to 2D dynamics
is larger than the term corresponding to the dynamics of the perpendicular
component. Hence we conclude that this particular type of heterochiral interaction
mainly contributes to the 2D dynamics.
\begin{figure}[tbp]
\begin{center}
\includegraphics[width=\columnwidth]{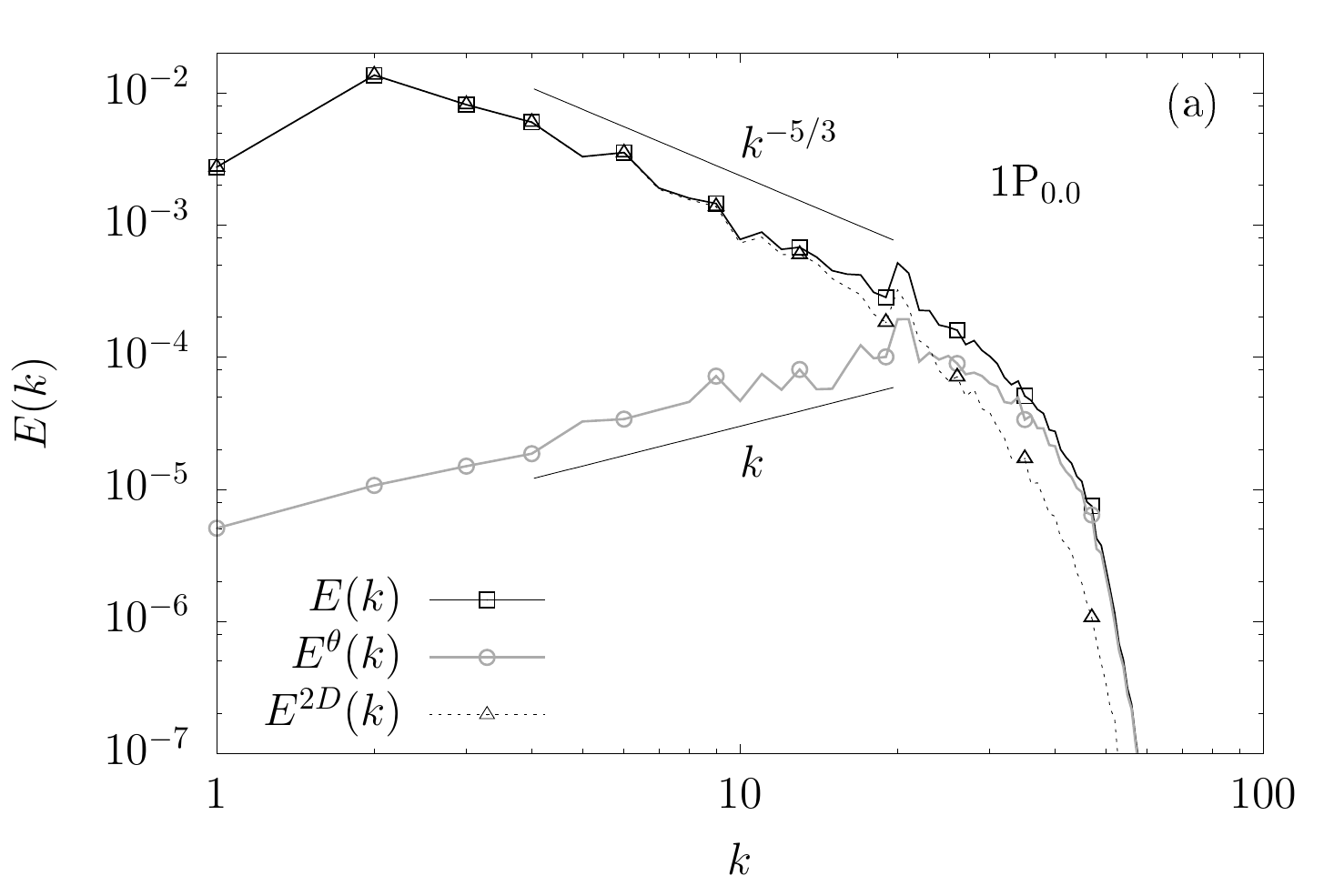} 
\includegraphics[width=\columnwidth]{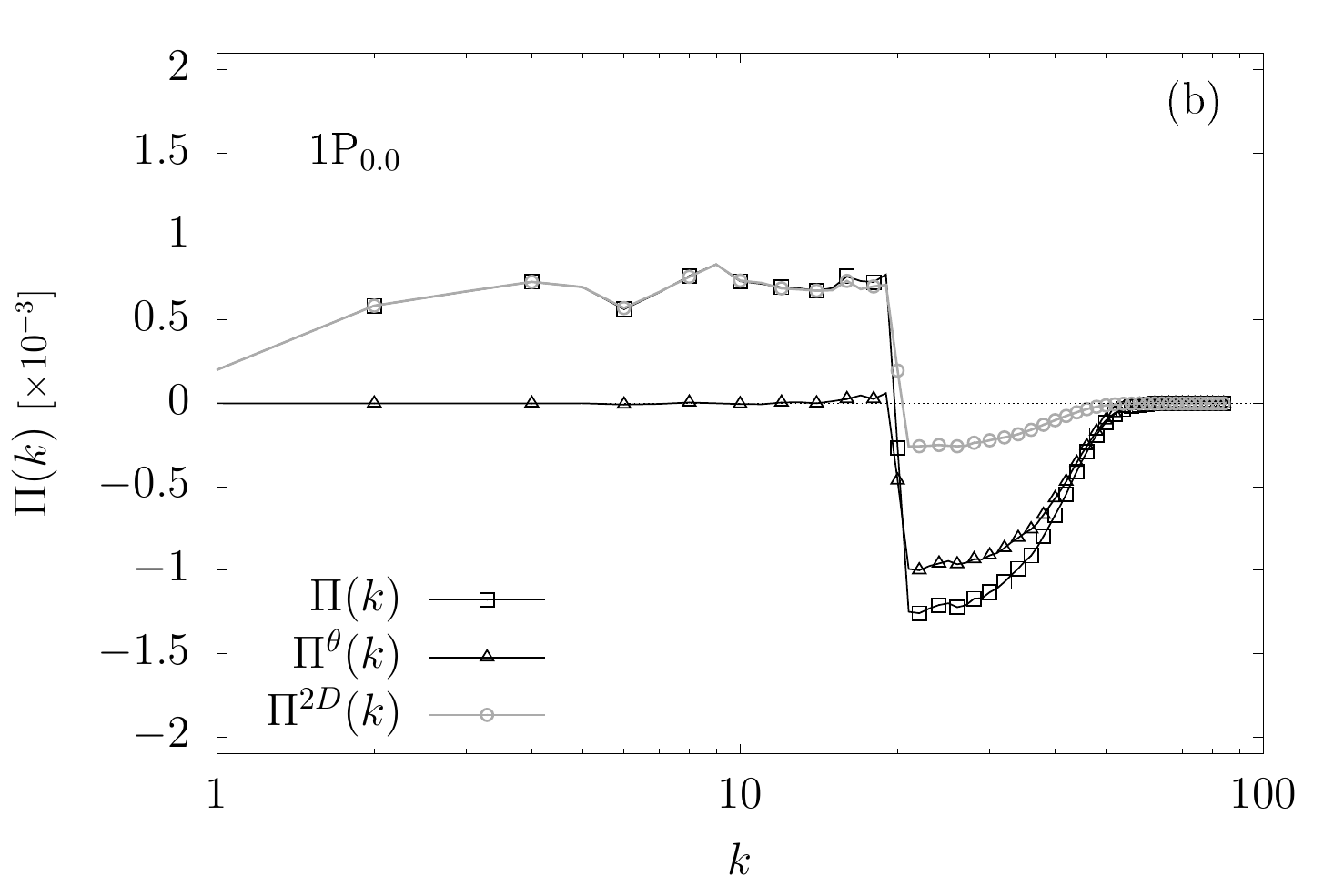} 
\includegraphics[width=\columnwidth]{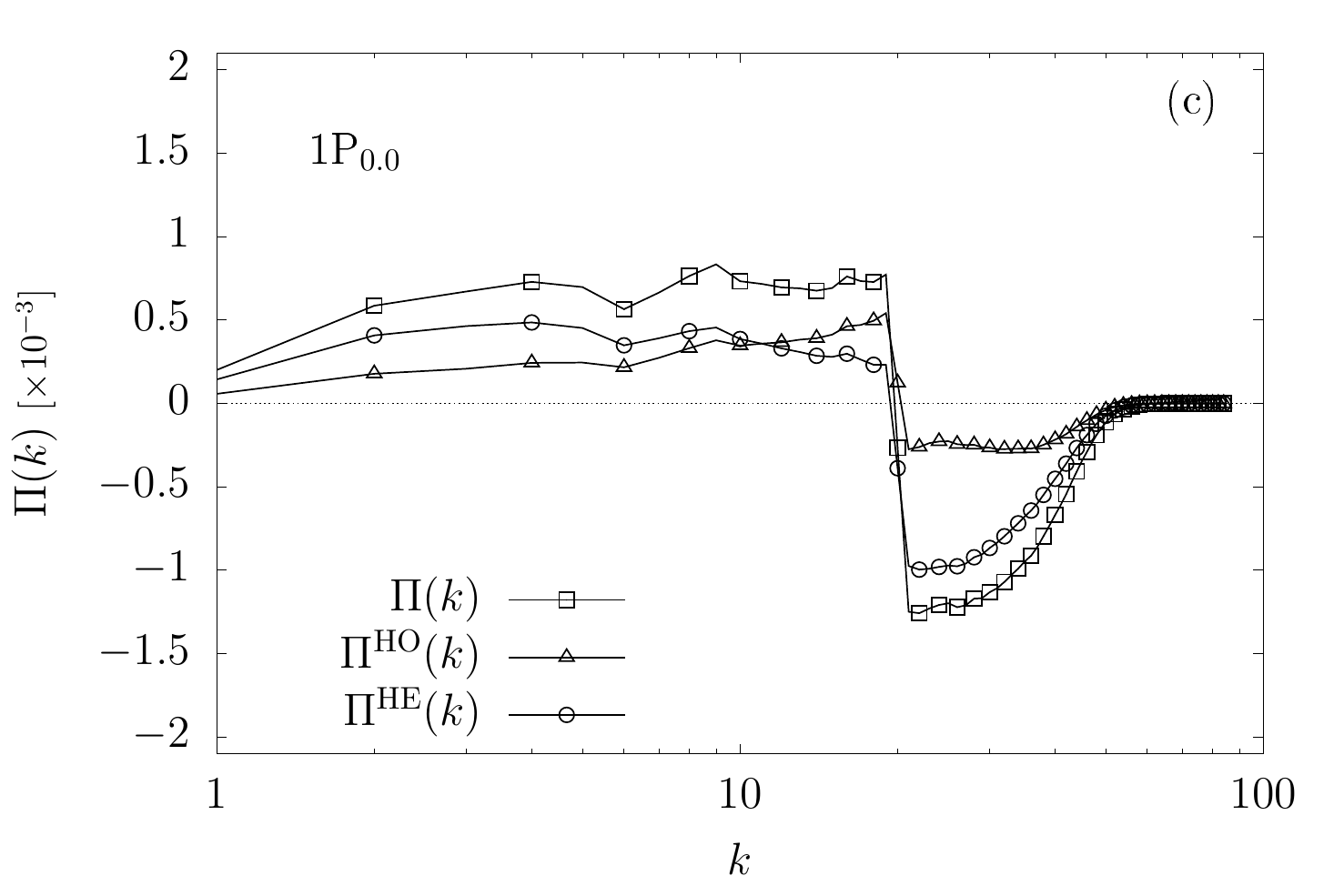} 
\caption{Decompositions of a single 2D3C flow 1P$_{0.0}$. (a) Energy spectra: $E(k)$, $E^{\text{2D}}(k)$ and $E^{\theta}(k)$.
         (b) Fluxes: $\Pi(k)$, $\Pi^{\text{2D}}(k)$ and $\Pi^{\theta}(k)$.
         (c) Fluxes: $\Pi(k)$, $\Pi^{\text{HO}}(k)$ and $\Pi^{\text{HE}}(k)$.}
\label{fig:1plane-flux-spectra}
\end{center}
\end{figure}

\subsubsection{Discussion}
In contrast to the fully 3D case, the coupling factor
$(p^2-q^2)p \sin\varphi_{k,p}$ {corresponding to the} 
2D evolution {(see also eq.~\eqref{eq:Fevolution_perp} in Appendix \ref{app:coupling_factors})} is
helicity-independent. This has two consequences: \\
(i) in 2D ($\theta = 0$) all couplings between the stream functions ({\em i.e.}~all helical interactions)
    are equally weighted for a given triad geometry, \\
(ii) a stability analysis based on single triads corresponding to 2D dynamics
extracted from Eq.~\eqref{eq:Fevolution_psi}
gives the same results for all helicity combinations as shown in Ref.~\onlinecite{Waleffe92}. \\ 
{In 2D} all interactions therefore lead to an inverse energy cascade and the coupling factor
is the same. 
That is, all classes of helical interactions produce an inverse energy 
cascade when restricted to the evolution in the plane.  
Concerning the evolution of the perpendicular component, 
the coupling factor $(s_qq-s_pp)kp \sin\varphi_{k,p}$ is helicity-dependent in such a way that
helical interactions leading to a forward cascade (Classes (III) and (IV)) 
are higher weighted than those leading to an inverse energy cascade or a mixed energy transfer (Classes (I) and (II)).
This confirms that a passive scalar in 2D turbulence should display a direct energy cascade, in accord with 
the known phenomenology of passive scalar advection in 2D \cite{Falkovich01}, see Appendix \ref{app:passive_scalar} for 
a further discussion of this point. 
\\

\noindent 
Figure \ref{fig:1plane-flux-spectra}(a-c) presents the energy spectra 
of the planar and perpendicular components 
\begin{align}
E^\theta(k)  &=  \frac{1}{2}\sum_{|{\bf k}|=k} |\hat\theta_\bk|^2 \ , \\
E^{2D}(k)  &=  \frac{1}{2}\sum_{|{\bf k}|=k} |\fbuperp_\bk|^2 \ ,
\end{align}
and the corresponding fluxes
\begin{align}
\Pi^\theta(k)  &=  -\sum_{k'=1}^{k}\sum_{|{\bk}|=k'} 
               \hat\theta_\bk \hspace{-1em} \sum_{\bk+\bp+\bq=0} \hspace{-1em} 
(i\bk \cdot \fbuperp_\bp) \hat\theta_\bq  \ , \\
\Pi^{2D}(k)  &= - \sum_{k'=1}^{k}\sum_{|{\bk}|=k'} 
               \fbuperp_\bk \cdot \hspace{-1em}\sum_{\bk+\bp+\bq=0}\hspace{-1em}
 (i\bk \cdot \fbuperp_\bp) \fbuperp_\bq\ .
\end{align}
of  the single 2D3C flow and the total, homo- and heterochiral energy fluxes 
\begin{align}
\Pi(k)  &= - \sum_{k'=1}^{k}\sum_{|{\bk}|=k'} 
               \hat \bu_\bk \cdot \hspace{-1em} \sum_{\bk+\bp+\bq=0} \hspace{-1em}(i\bk \cdot \hat \bu_\bp) \hat\bu_\bq\ , \\
\Pi^{\rm HO}(k)  &=  -\sum_{k'=1}^{k}\sum_{|{\bk}|=k'} \sum_{s\in\{+,-\}} 
\hspace{-1em} \hat \bu^s_\bk \cdot \hspace{-1em} \sum_{\bk+\bp+\bq=0}\hspace{-1em} 
 (i\bk \cdot \hat \bu^s_\bp) \hat\bu^s_\bq\ , \\
\Pi^{\rm HE}(k) &= \Pi-\Pi^{\rm HO}(k) \ ,
\end{align}
respectively.
The data has been obtained by DNS of a 2D3C flow as described in detail in
Sec.~\ref{sec:simulations}.  From Fig.~\ref{fig:1plane-flux-spectra}(a) we
observe clearly the predicted behavior of the 2D and the perpendicular
component with $E^{2D}(k)$ displaying an inverse-cascade $k^{-5/3}$-scaling
while $E_\theta(k)$ shows equipartition scaling in the same wavenumber range.
As a result, the total energy spectrum $E(k)= E^{2D}(k) + E^\theta(k)$ at $k
\sim O(1)$ is mainly given by the 2D component.  The separate fluxes
$\Pi^{2D}(k)$ and $\Pi^\theta(k)$ are shown in
Fig.~\ref{fig:1plane-flux-spectra}(b), which confirms that the inverse energy
flux is exclusively generated by the 2D dynamics, while the perpendicular
component $\theta$ has zero inverse flux in agreement with the 2D absolute
equilibrium scaling of its energy spectrum shown in
Fig.~\ref{fig:1plane-flux-spectra}(a).  Interestingly enough, concerning the
energy flux in the helical decomposition, the inverse cascade of total energy
is given to a large extent by the homochiral fluxes according to
Fig.~\ref{fig:1plane-flux-spectra}(c), which is consistent with the analysis
presented in Sec.~\ref{sec:cascade_directions}, since homochiral interactions
mainly contribute to the evolution in the plane. However, the homochiral flux
is not reproducing the total energy flux entirely and heterochiral interactions
also contribute to the inverse flux at all scales larger than the forcing
scale, as expected for 2D dynamics which must be helicity-insensitive. A more
remarkable difference between homo- and heterochiral components of the flux is
detectable for scales smaller than the forcing scale, where the homochiral
contribution is fully negligible.  
{
Notice that $\Pi^{\rm HO}(k)$  accounts for nearly one-third to one-half
of $\Pi(k)$ for $k < k_f$ (see Fig.~\ref{fig:1plane-flux-spectra}(c)), 
which according to Fig.~\ref{fig:1plane-flux-spectra}(b)
is given by the 2D dynamics only. However, the geometry of the nonlinear
coupling in the plane suggests that 
the homo- and heterochiral contributions to the planar dynamics are identical, 
and one would expect $\Pi^{\rm HO}(k)\simeq \Pi^{2D}(k)/4$ if the 
dynamics was purely 2D. 
However, the heterochiral contribution to $\Pi(k)$ is mostly forward 
(in the wavenumber range $k > k_f$), hence we expect less energy to be transferred
upscale by heterochiral interactions compared to homochiral interactions.    
In summary, homochiral interactions enhance the 2D physics and 
lead to a better inverse cascade in 2D3C flows than heterochiral interactions.
}
\\

\section{Numerical Simulations} \label{sec:simulations}
The analytical results shed some light on the fundamental 
properties of the 2D3C Navier-Stokes equations by disentangling the dynamics of the 
plane from that of the perpendicular component, and they provided some qualitative results
concerning the particular dynamics of homochiral interactions. 
However, in order to study their effective importance for realistic 3D or quasi 3D flows, we need to perturb the basic 2D3C flows.
We do it in different steps. First we start by coupling three 2D3C flows each one restricted on one of the three perpendicular planes $(x,y)$, $(y,z)$ and $(x,z)$.
The resulting dynamics is not a simple superposition of the three 2D3C dynamics because there will be triads that couple the three planes. Hence, we expect to have a mixture of 2D and 3D
phenomenology, depending on the relative weights of  triads fully contained in each one of the three planes and triads with vertices on at least two different planes. 
{In particular, the additional inviscid invariants specific to single 2D3C flows 
are no longer conserved, and  
the only inviscid invariants of the coupled 2D3C flows are those of the full 
3D Navier-Stokes equations, {\em i.e.}, the total energy and the kinetic helicity.
Furthermore, the splitting into two stream functions is not possible any longer.
} 
%
\\
We carry out two series of simulations, series {\bf 1P} refers to the base flow being a single 2D3C flow  
in the $(x,y)$-plane while series {\bf 3P} consist of three 2D3C flows as the base configuration.\\
Furthermore, we will also successively decorate both sets of base flow by adding randomly (but quenched in time) modes that are fully 3D, 
{\em i.e.}~taken blindly in the 3D Fourier space.
By changing the percentage $\alpha$ to have $3D$ modes we will be able to move from $\alpha=0$ where we have either the {\bf 1P} or {\bf 3P} configuration to a fully resolved 3D Navier-Stokes case when $\alpha=1$. 
We  will mainly look for a change in the cascade direction from inverse (2D) to 
direct (3D), leaving to further studies other important issues as, e.g., the impact on small-scale intermittency.
 In order to do that we always force the system at small scales and to achieve a large scale separation while still resolving some small-scale turbulence,
we use higher order  hyperviscous Navier-Stokes 
equations   
\begin{align}
\label{eq:nse}
&\partial_t \bu = - \nabla \cdot (\bu \otimes \bu) - \nabla P + \nu (-1)^{n + 1}\Delta^{n}\bu + \bF \ , \\
\label{eq:incomp}
&\nabla \cdot \bu = 0 \ ,
\end{align}
where $\bv$ denotes the velocity field, $P$ the pressure, $\nu$ the 
(hyper)viscosity, $\bF$ an external force and $n=4$  the power of the Laplacian. 
Equations \eqref{eq:nse}-\eqref{eq:incomp} are stepped forwards in time using 
a pseudospectral code with full dealiasing according to the two-thirds rule \cite{Patterson71} 
on $256^3$ collocation points in a triply periodic domain $V$ of size $L=2\pi$, such that the
smallest resolved wavenumber is $k_{\rm min} = 1$ and the largest resolved 
wavenumber is $k_{\rm max} = 85$. 
The external force $\bF$ is given by a $\delta(t)$-correlated random process
in Fourier space 
\be
\langle \hat \bF_\bk(t) \hat \bF^*_\bq(t') \rangle = F \delta_{\bk,\bq} \delta(t-t')
\hat Q_\bk,
\ee
where $\hat Q_\bk$ is a projector applied to guarantee incompressibility  
and $F$ is nonzero in a given band of Fourier modes concentrated at intermediate to small scales
$k_f \in [20,21]$.
The magnitude of the forcing $F= 0.15$ and the value of the 
hyperviscosity $\nu = 1.8 \times 10^{-13}$ are the same for all simulations. 
Depending on the type of simulation, Eqs.~\eqref{eq:nse}-\eqref{eq:incomp} are Galerkin projected on the
appropriate subspaces, {\em i.e.}~the $(x,y)$-plane in isolation or the $(x,y)$, $(y,z)$ and 
$(x,z)$-planes only. The same can be done when we randomly add modes in the whole Fourier space. 
 This is achieved  
through a probabilistic projector $\mathfrak{P}_\alpha$ acting on the velocity field \cite{Frisch12,Lanotte15,Buzzicotti16} in the volume outside 
the planes
\be
\bu_\alpha (\bx,t) \equiv 
\mathfrak{P}_\alpha \bu(\bx,t) = \sum_{\bk \in \mathbb{Z}}\gamma_\bk \hat \bu_\bk(t) e^{i\bk\cdot \bx} \ ,
\ee
where 
\be
\gamma_\bk = 
\begin{cases}
1 \quad \text{with probability } \alpha \ , \nonumber \\ 
0 \quad \text{with probability } (1-\alpha) \ , 
\end{cases}
\ee
for $0 \leqslant \alpha \leqslant 1$.
The projector $\mathfrak{P}_\alpha$ is determined at the start of the simulation and remains unchanged
afterwards. In order to guarantee that the evolution of the projected field $\bu_\alpha$ remains in the same 
subspace $V_\alpha \equiv \mathfrak{P}_\alpha(V)$, the nonlinear term in the Navier-Stokes equations must be 
re-projected at each iteration step.
\begin{table}
\begin{center}
\begin{tabular}{cccccccc}
   & 1P$_{0.0}$ & 1P$_{0.01}$ & 1P$_{0.05}$ & 1P$_{0.1}$ & 1P$_{0.15}$ & 1P$_{0.2}$ & 1P$_{0.3}$  \\
  \hline
   $\alpha$ & 0.0 & 0.01 & 0.05 & 0.1 & 0.15 & 0.2 & 0.3\\
   $\varepsilon$ \tiny{$[\times10^{-3}]$} &  1.3 & 1.3 & 2 & 2.7 & 2.5 & 3 & 3.2 \\
   $U$ & 0.42 & 0.42 & 0.33 & 0.2 & 0.18 & 0.18 & 0.17\\
   $\ell$ &  1.2 & 1.2 & 0.7 & 0.18 & 0.15 & 0.132 & 0.128 \\
   $t/T_f$ &  90 & 90 & 90 & 90 & 90 & 90 & 90\\
  \hline
\vspace{0.3em} \\
  \hline
   & 3P$_{0.0}$ & 3P$_{0.01}$ & 3P$_{0.05}$ & 3P$_{0.1}$ & 3P$_{0.15}$ & 3P$_{0.2}$ & 3P$_{0.3}$  \\
  \hline
   $\alpha$ & 0.0 & 0.01 & 0.05 & 0.1 & - & - & 0.3 \\
   $\varepsilon$ \tiny{$[\times10^{-3}]$} & 6 & 6 & 6 & 6 & - & - & 6\\
   $U$ & 0.41 & 0.36 & 0.27 & 0.24 & - & - & 0.2 \\
   $\ell$ &  0.45 & 0.33 & 0.15 & 0.14 & - & - & 0.128 \\
   $t/T_f$ &  90 & 90 & 45 & 45 & - & - & 45 \\
  \hline
  \end{tabular}
\caption{
Specifications of the numerical simulations. 
Series 1P corresponds to a single 2D3C base flow 
and series 3P to a three coupled 2D3C base flows. 
The fraction of added 3D modes is denoted by $\alpha$, 
$\varepsilon$ is the dissipation rate,
$U=\sqrt{2E}$ the root-mean-square velocity, 
$\ell=(\pi/2U^2)\int dk \ E(k)/k$ the integral scale, 
and $t/T_f$ the run time in units of 
forcing-scale eddy turnover time 
$T_f=(2\pi/(Fk_f))^{1/2}$. 
The values for $U$, $\ell$ and $\eps$ are time averages for 
runs reaching stationary state and otherwise correspond to 
values taken at the end of the simulations. 
}
\end{center}
\label{tbl:simulations}
\end{table}
A summary of specifications of the simulations including the chosen values of $\alpha$ 
is provided in table I. 
In the following we will use the short-hand notation
$3P_\alpha$ and $1P_\alpha$ to indicate simulations starting from a basic 1P or 3P configuration with a given value
of $\alpha$. 
  
\subsection{3 Coupled 2D3C-flows}
\begin{figure*}[t]
\begin{center}
\hspace{-0.2cm}
\includegraphics[width=\columnwidth]{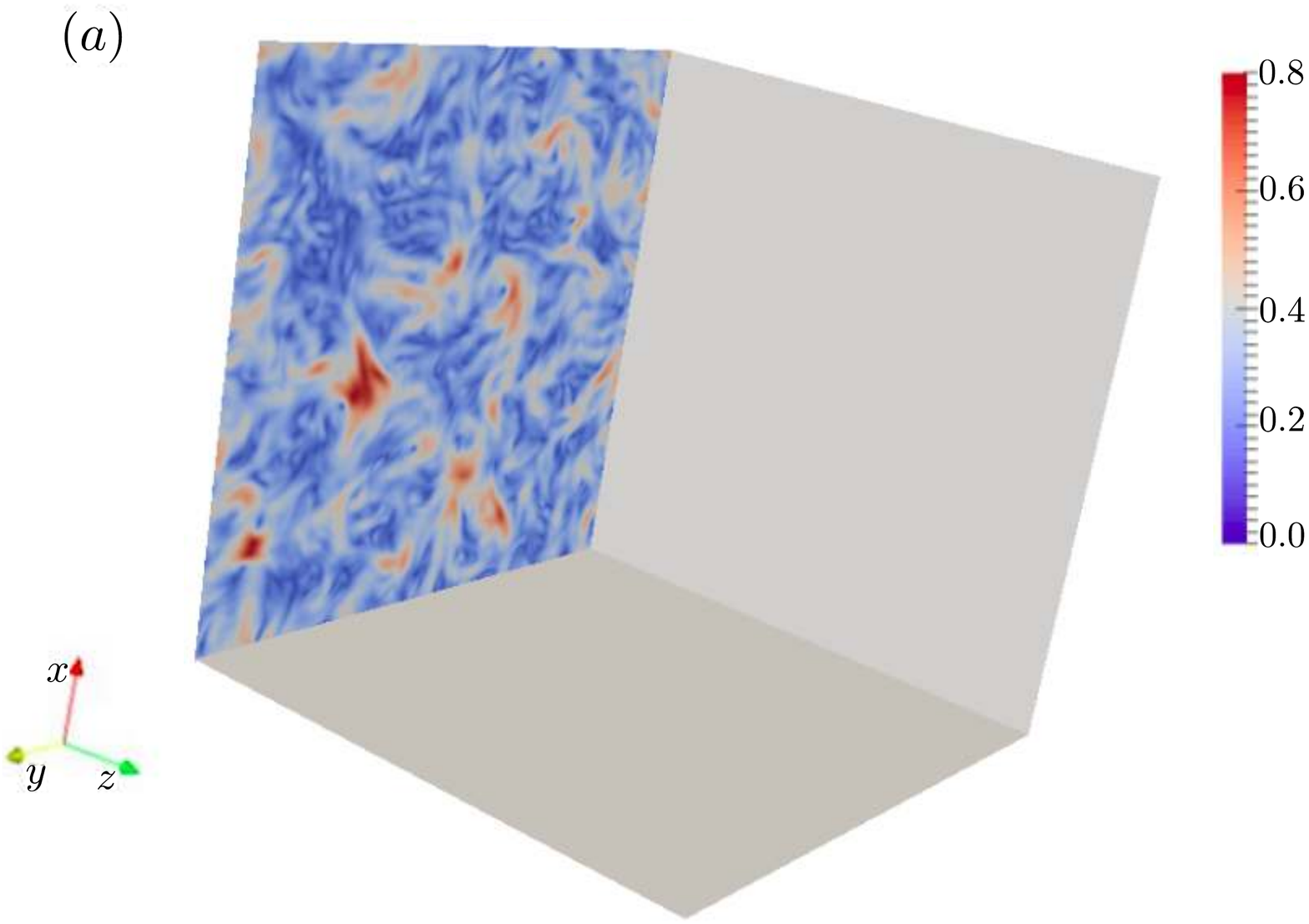}
\includegraphics[width=\columnwidth]{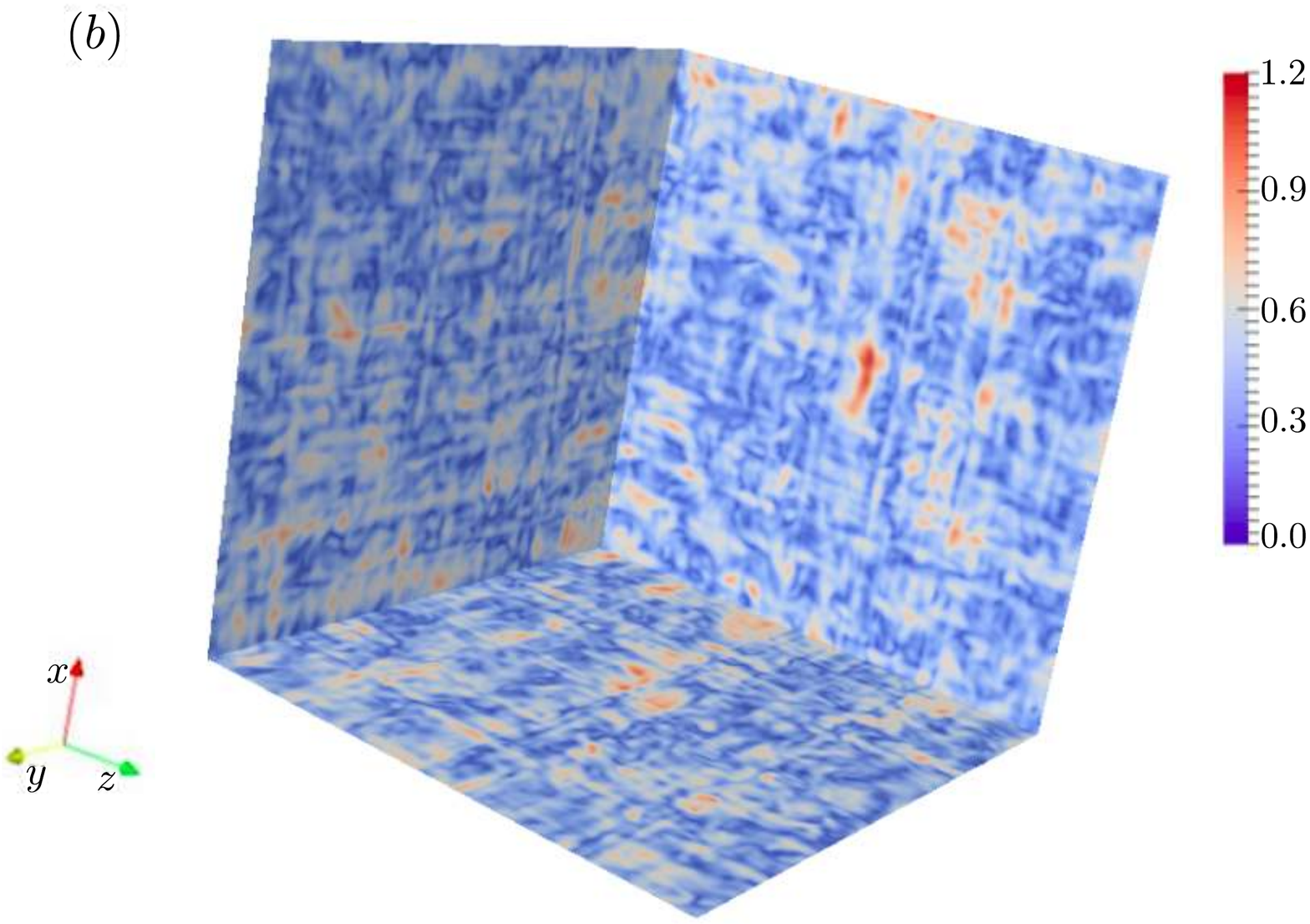}
\caption{Visualizations of \obs{the kinetic energy of} a 2D3C flow 1P$_{0.0}$ (a) and three coupled 2D3C flows 3P$_{0.0}$ (b).
The characteristic cross-pattern visible in panel (b) results from the 
coupling of the respective perpendicular components of the single 2D3C flows
in the three planes.
}
\label{fig:visualisations_1plane_3d}
\end{center}
\end{figure*}
\begin{figure}[tbp]
\begin{center}
\includegraphics[scale=0.5]{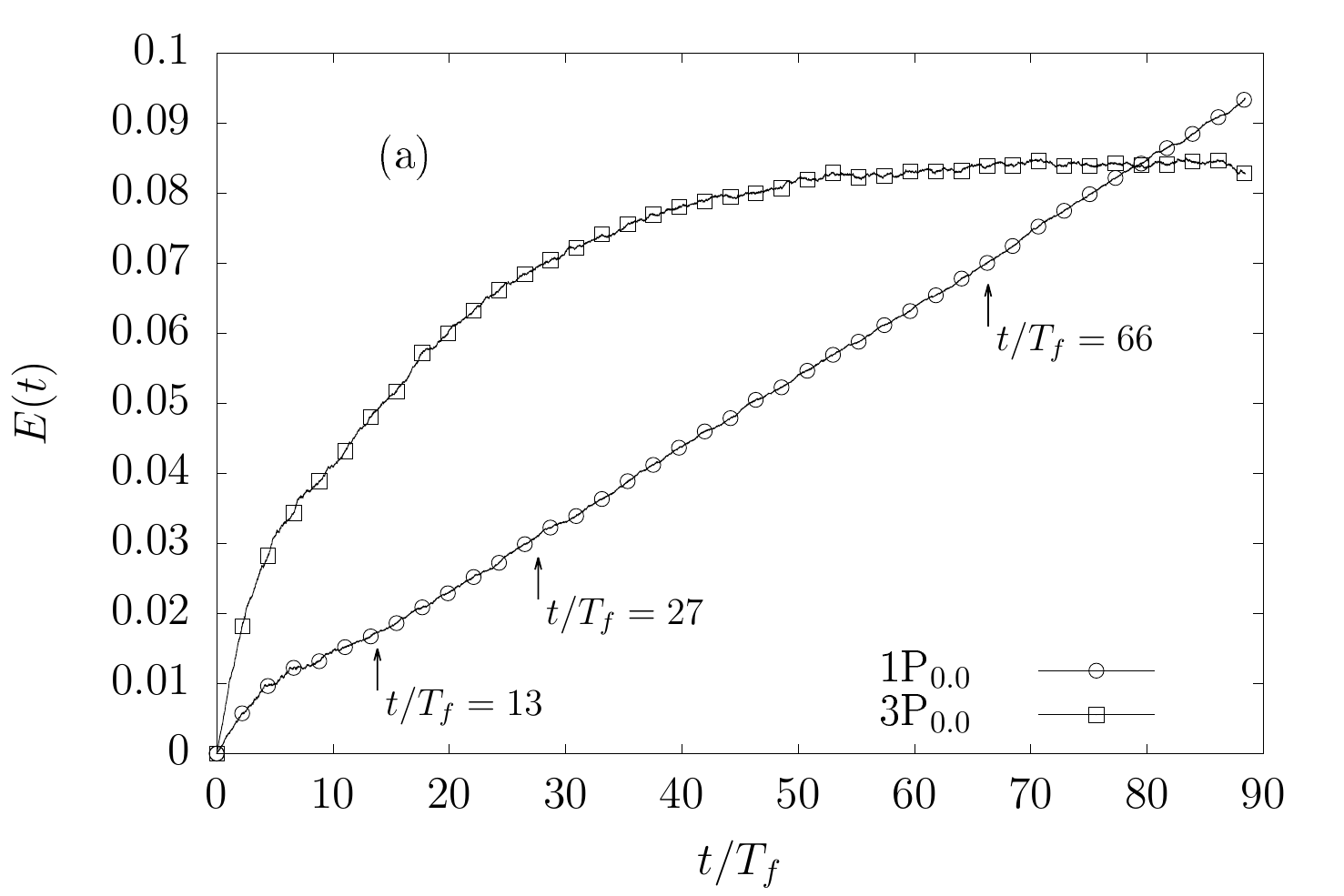} 
\includegraphics[scale=0.5]{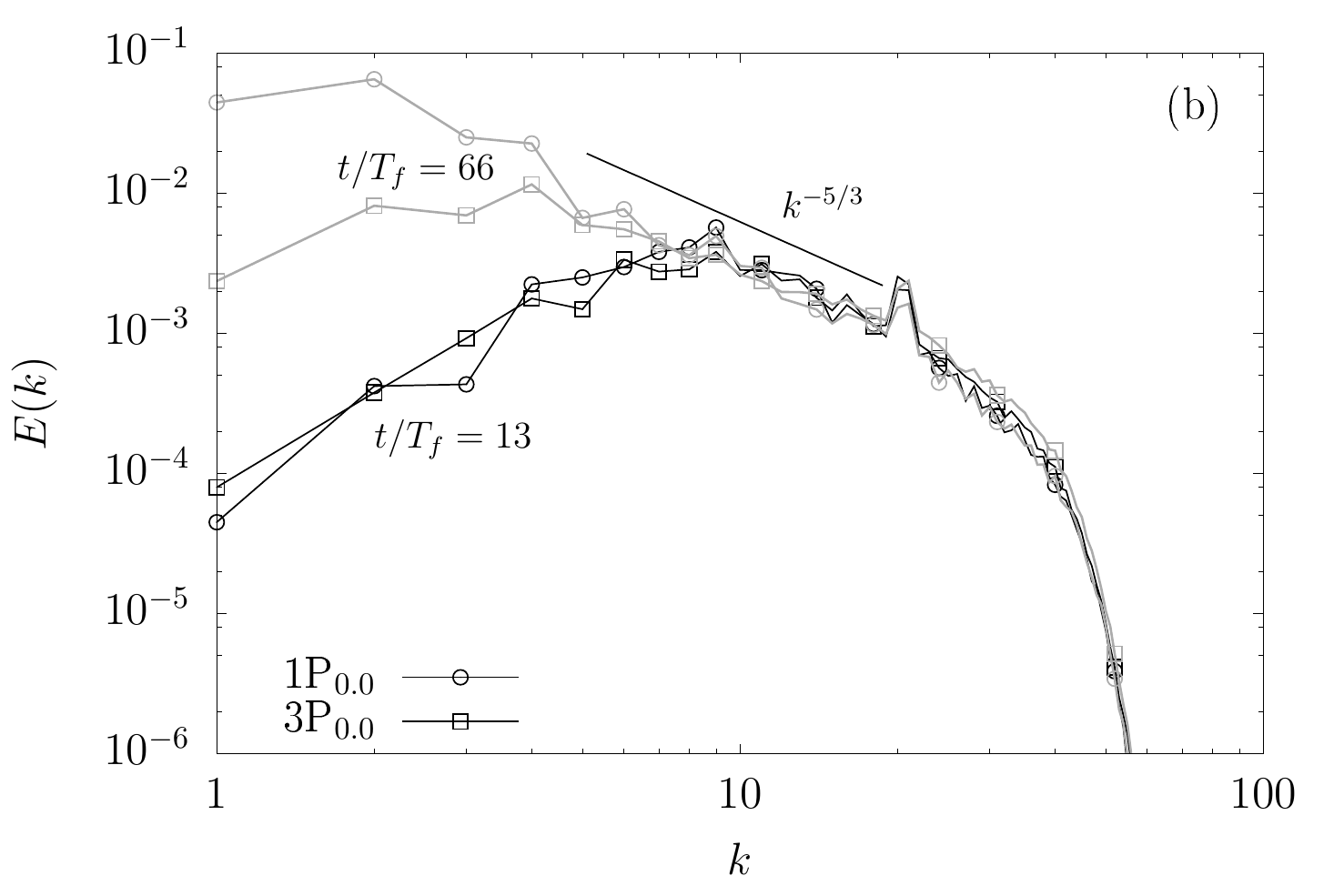} 
\caption{
(a) $E(t)$ temporal evolution for single 1P$_{0.0}$ (open circles) and three coupled 2D3C flows, 3P$_{0.0}$ (open squares).
(b) $E(k)$ for 1P$_{0.0}$ (open circles) and 3P$_{0.0}$ (open squares), measured at time $t/T_f=13$ (black lines) and $t/T_f=66$ (gray lines) during the evolution.
}
\label{fig:evolution_alpha0}
\end{center}
\end{figure}
Before discussing the effect of additional 3D modes, we describe the evolution
of the two base configurations $1P_{0.0}$ and $3P_{0.0}$, that is, a single 2D3C flow
and three coupled 2D3C flows.  The single 2D3C flow is expected to display an
inverse cascade of the 2D component while the third component evolves as a
passive scalar.  The coupled 2D3C flows should also display an inverse cascade
corresponding to decoupled or weakly coupled 2D3C dynamics. However, although
the three planes may be weakly coupled at small scales, at the large scales, {\em i.e.}~small $k$,  the
coupling will become more significant due to a larger relative fraction of triads coupling wavevectors of the three planes. 
Hence,  their coupling  should produce 3D dynamics and the inverse cascade is
expected to stop, leading to
{a transient inverse energy transfer which does not lead to the formation of a large-scale condensate.}
This
is exactly the opposite of what typically happens in geophysical flows, where the small-scale dynamics is almost \obs{3}D and only the large scale
evolution is feeling the 2D confinement. Our three-plane 2D3C configuration has a reverted behavior.
\\
\noindent
The time evolution of the total energy for configurations $1P_{0.0}$ and $3P_{0.0}$ is shown in Fig.~\ref{fig:evolution_alpha0}(a), 
and we clearly observe that the kinetic energy of the single 2D3C flow grows linearly, which is a tell-tale
sign of an inverse energy cascade, 
while the coupled 2D3C dynamics 
results eventually in the formation of a stationary state as expected. 
Figure \ref{fig:evolution_alpha0}(b) shows the total energy spectra 
corresponding to snapshots at $t/T_f=13$ and $t/T_f=66$ in the time evolution of both
configurations corresponding to the arrows in
Fig.~\ref{fig:evolution_alpha0}(a). As can be seen, both configurations show an
inverse cascade $k^{-5/3}$-scaling at $k<k_f$, which it stops around $k=10$ for
run $3P_{0.0}$ while it continues to larger scales for run $1P_{0.0}$, as
expected.  Visualizations of the two base configurations are shown in
Fig.~\ref{fig:visualisations_1plane_3d}, where it can be seen that the single
2D3C dynamics results in the formation of large-scale structures
while that of
the coupled 2D3C dynamics does not.  \\

\begin{figure}[tbp]
\begin{center}
\includegraphics[width=\columnwidth]{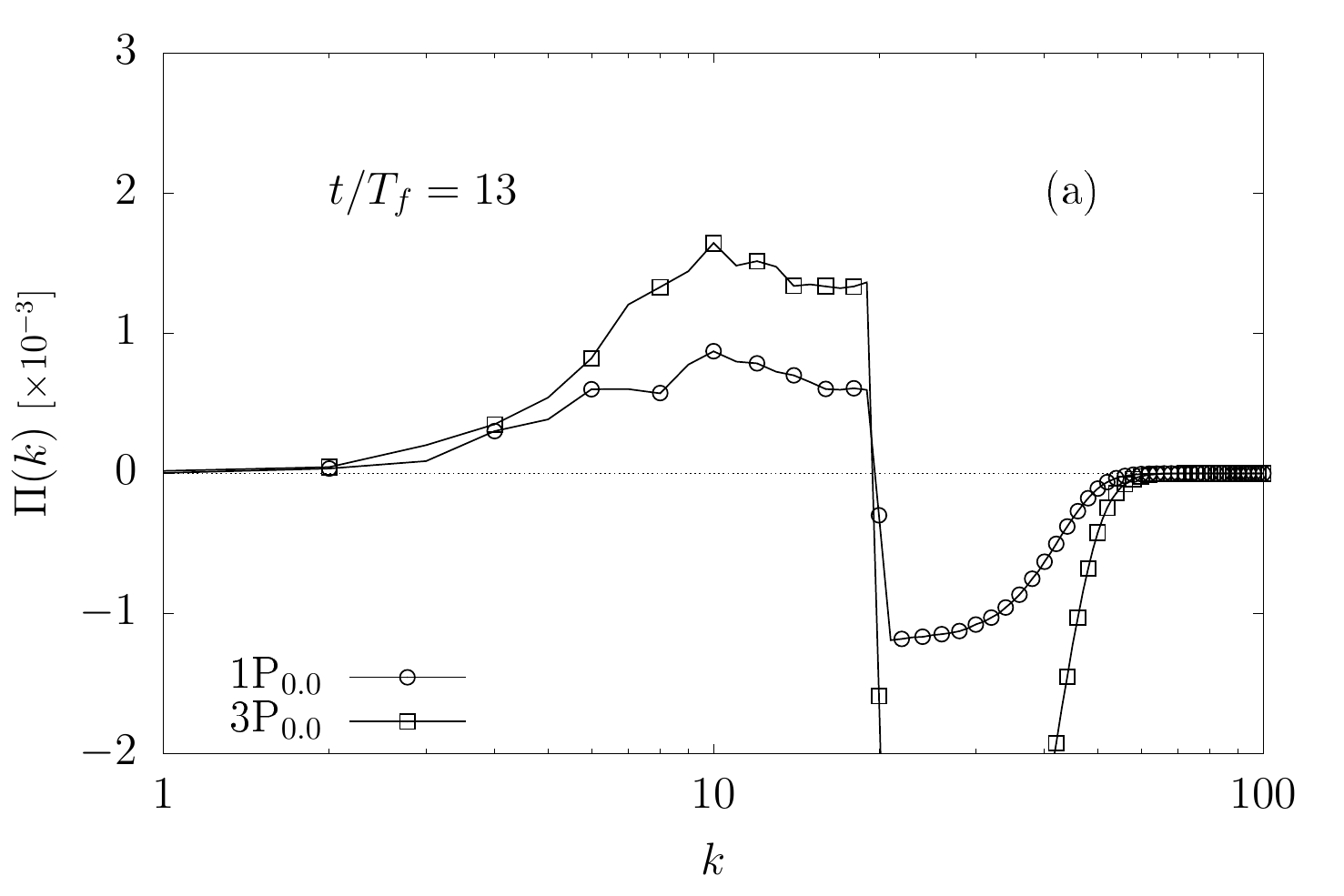} 
\includegraphics[width=\columnwidth]{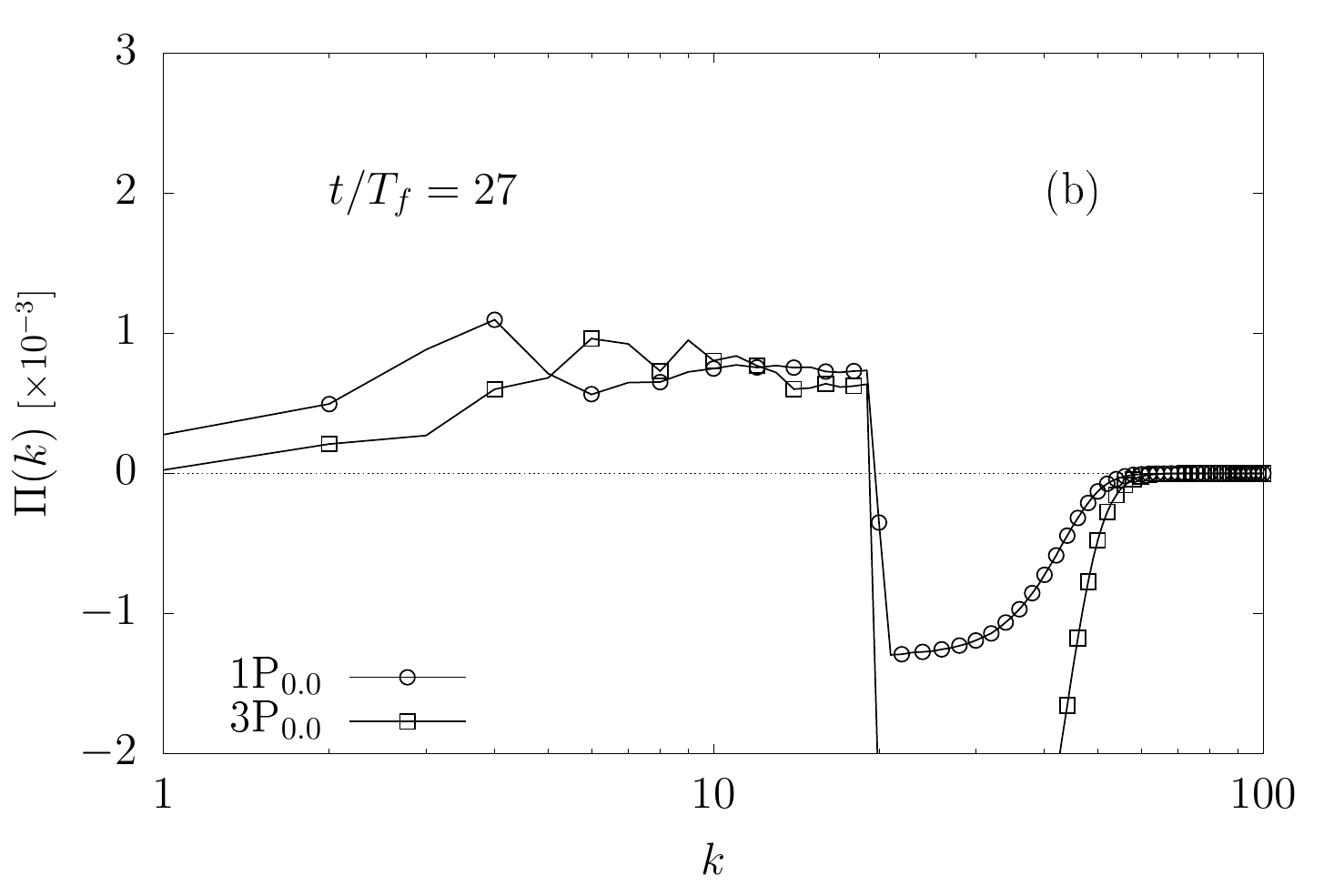} 
\includegraphics[width=\columnwidth]{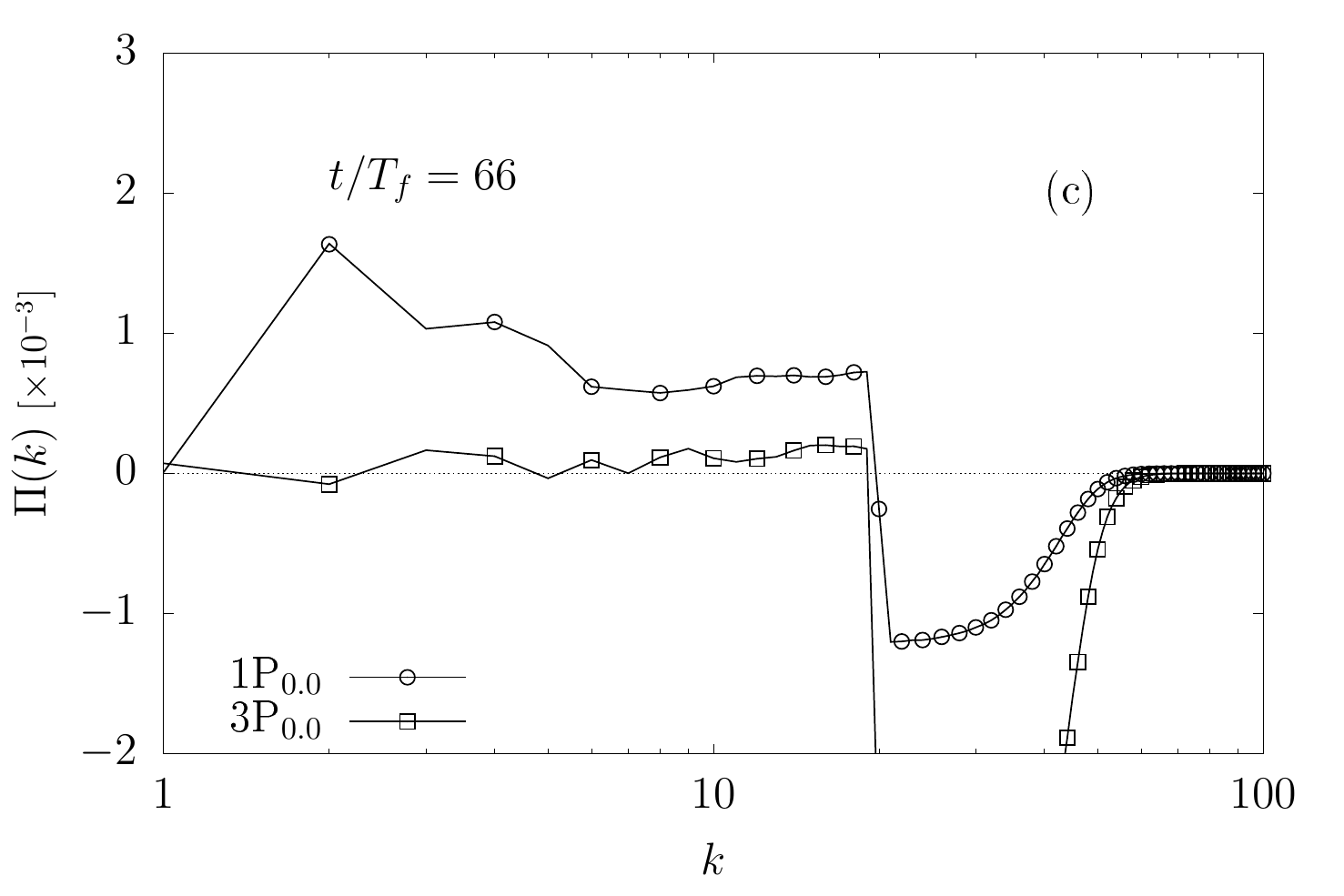} 
\caption{$\Pi(k)$ for a single 1P$_{0.0}$ (open circles) and three coupled 2D3C flows 3P$_{0.0}$ (open squares) at $t/T_f=13$ (a), 
$t/T_f=27$ (b), $t/T_f=66$ (c).
}
\label{fig:fluxes_alpha0}
\end{center}
\end{figure}

\noindent
The energy fluxes $\Pi(k)$ for both configurations are shown in
Figs.~\ref{fig:fluxes_alpha0}(a-c) at $t/T_f=13$, $t/T_f=27$ and $t/T_f=66$, respectively.
At early times in the evolution ($t/T_f=13$) we observe a clear inverse energy flux
in both runs. As can be seen by comparison of the three figures, the inverse
flux remains for run $1P_{0.0}$, while it diminishes as run $3P_{0.0}$ approaches the
stationary state. Figure ~\ref{fig:fluxes_alpha0}(c) corresponds to a snapshot
in time after saturation of run $3P_{0.0}$, and we observe that the inverse flux
corresponding to run $3P_{0.0}$ now nearly vanishes. In the present setup this does
not imply that no energy is transferred into the large scales, since the
corresponding energy spectrum shown in Fig.~\ref{fig:evolution_alpha0}(b) has
not transitioned to an equipartition spectrum as would be the case for a fully
3D flow subject to small-scale forcing \cite{Dallasetal15}. As such, the absence
of a pronounced inverse flux must result from a balance between inverse and
forward fluxes leading to a stationary state and it cannot be interpreted as a
sign of a transition to fully 3D dynamics. The latter point is further
discussed in the context of the contribution of homo- and heterochiral
contributions to the total flux in Sec.~\ref{sec:homochiral_numerics}.      
\\

\subsection{2D-3D transition} \label{sec:2d3d-transition}
\begin{figure}[t]
\begin{center}
\includegraphics[width=\columnwidth]{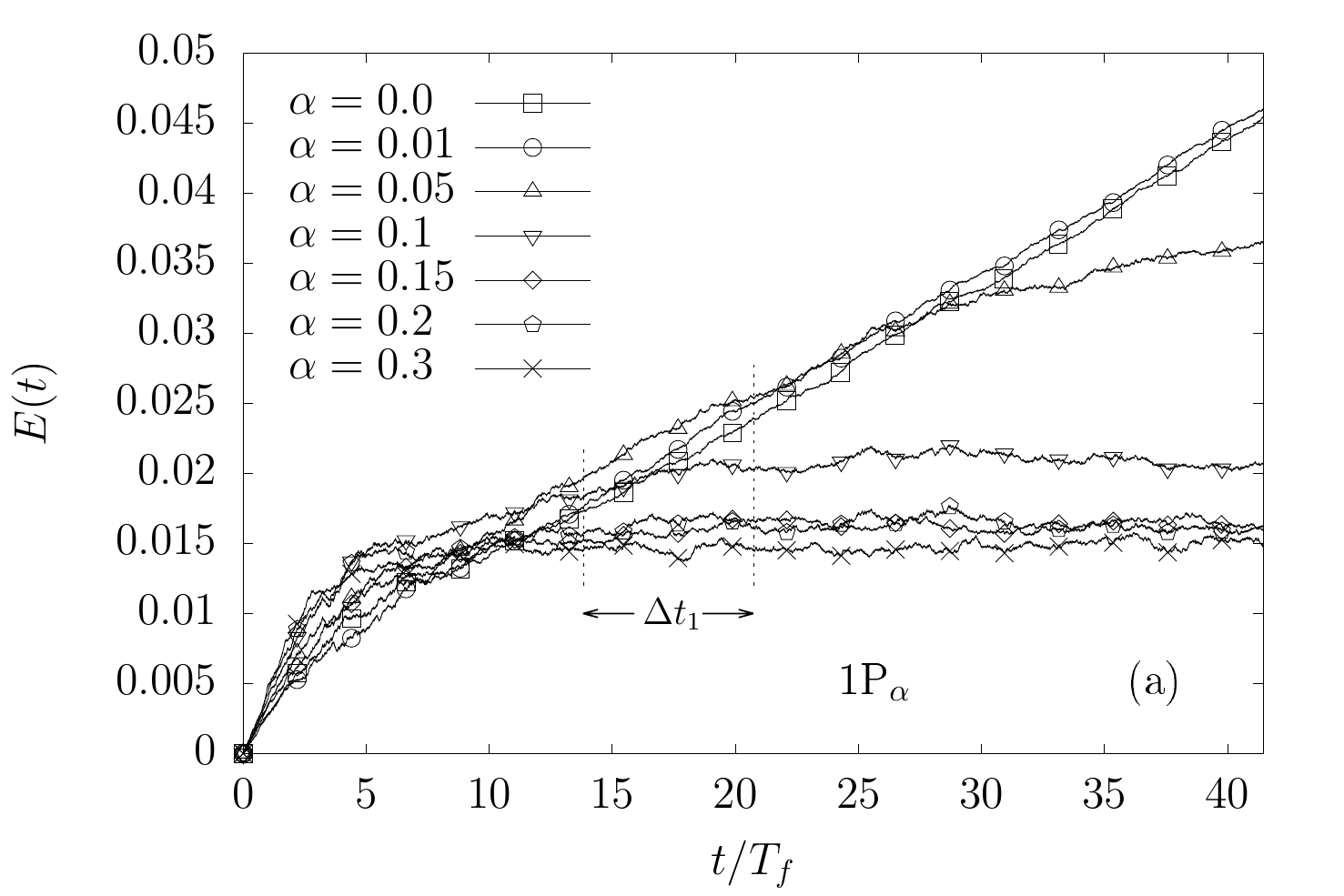} 
\includegraphics[width=\columnwidth]{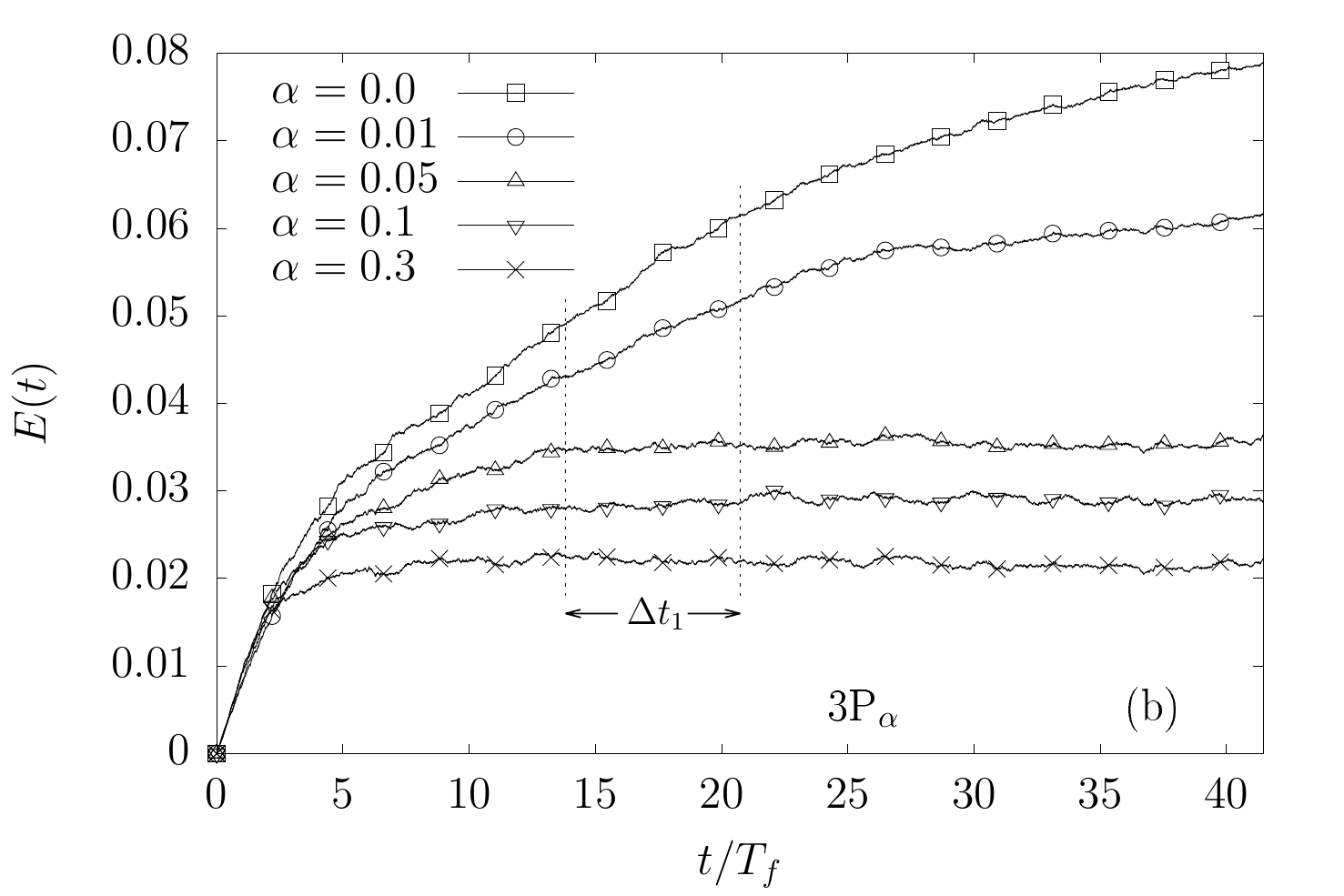} 
\caption{$E(t)$ as a function of time for different $\alpha$.
(a) Single 2D3C flow 1P$_{\alpha}$. (b) Three coupled 2D3C flows 3P$_{\alpha}$.
}
\label{fig:23planes_alpha_evol}
\end{center}
\end{figure}

Having described the main features of the base configurations 
$1P_{0.0}$ and $3P_{0.0}$, we now proceed to an investigation of the transition 
to fully 3D dynamics in both configurations.  
The time evolution of the total energy for each run in the two DNS series 1P and 3P corresponding to different 
percentage $0\leqslant \alpha \leqslant 0.3$ of added 3D Fourier
modes is presented in Fig.~\ref{fig:23planes_alpha_evol}(a) and (b), 
respectively. For both series we observe that by increasing $\alpha$ a stationary 
state is reached earlier in time and with a lower total energy. Furthermore, in both cases
no significant difference in the evolution of the total energy can be seen already for 
$\alpha =0.2$ and larger.
For series $1P_\alpha$ the results shown in Fig.~\ref{fig:23planes_alpha_evol}(a) 
suggest that the transition from 2D to 3D dynamics appears to occur at 
$\alpha < 0.2$, as run $1P_{0.2}$ does not show any transient inverse transfer 
while run $1P_{0.1}$ shows initially an inverse energy transfer 
which saturates around $t/T_f=17$. 
In contrast, the transition occurs 
faster as a function of $\alpha$ for series 3P, as run $3P_{0.05}$ reaches a stationary 
state at $t/T_f=13$ already and run $3P_{0.1}$ does not display an inverse energy transfer 
at all.
\\

\begin{figure}[tbp]
\begin{center}
\includegraphics[width=\columnwidth]{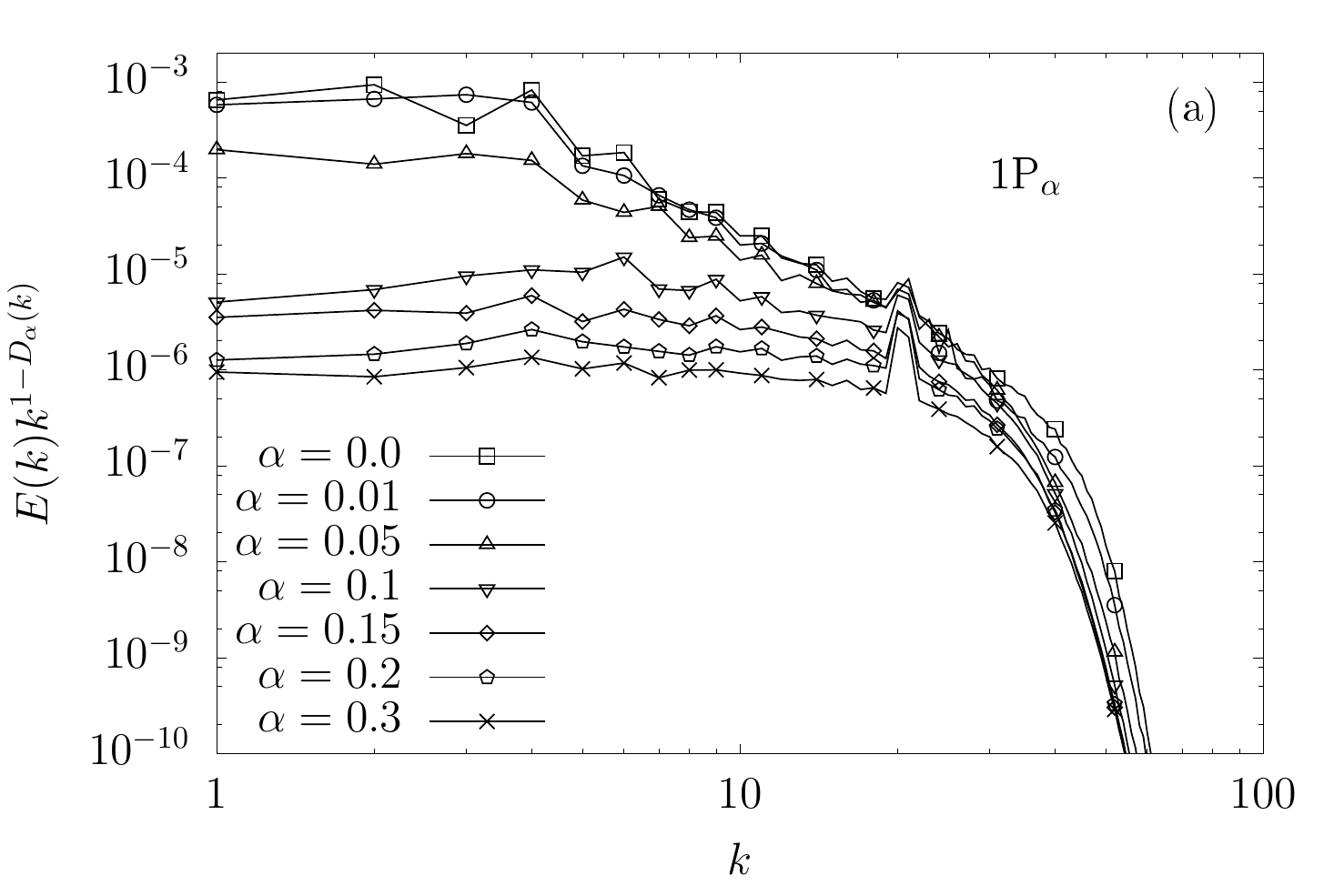} 
\includegraphics[width=\columnwidth]{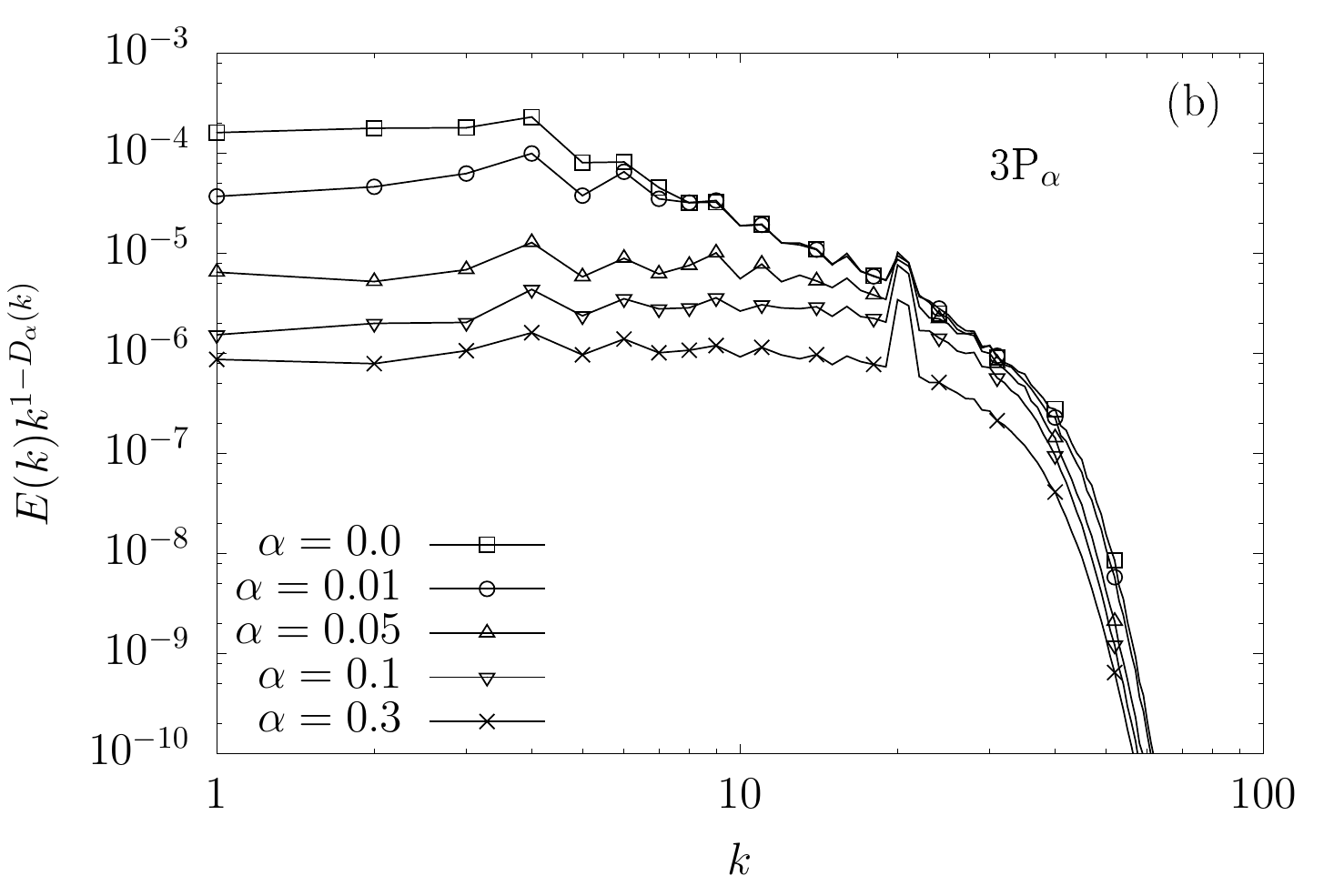} 
\caption{
$E(k)$ for different $\alpha$ compensated with the respective
absolute equilibrium prediction. (a) Single 2D3C flow 1P$_{\alpha}$, (b) three coupled 2D3C flows 3P$_{\alpha}$.
}
\label{fig:23planes_alpha_spect_comp}
\end{center}
\end{figure}
\noindent
These results are further substantiated by measurements of the energy spectra 
for the two series of DNSs obtained either during stationary state where applicable, or otherwise 
by time averaging over a short interval at late times in the simulation. 
The shape of the energy spectrum at wavenumber $k < k_f$ depends on the dimensionality 
of the flow and the direction of the energy flux.
In particular, in fully 3D dynamics the absence of an inverse total energy 
flux results in the Fourier modes $\bu_\bk$ in the range $k < k_f$ being in statistical equilibrium 
with equally distributed kinetic energy amongst them \cite{Hopf52,Kraichnan73,Dallasetal15}. 
Any residual inverse energy flux will perturb this equilibrium state
and hence will result in deviations from the expected scaling of the 
energy spectra. 
Figures \ref{fig:23planes_alpha_spect_comp}(a) and (b) show the energy spectra
for series 1P and 3P, respectively, compensated with the absolute equilibrium
prediction corresponding to the fraction of added 3D modes
\be
E_{\alpha}^{equil}(k) \sim k^{D_\alpha(k) -1} \ ,
\ee
where $D_\alpha(k) = 3$ for $\alpha = 1$, $D_\alpha(k) = 2$ for $\alpha = 0$, while for $\alpha \neq 0,1$
the value of $D_\alpha(k)$ is determined numerically by brute force  counting the
number of active modes in a given wavenumber shell. 
As can be seen from Fig.~\ref{fig:23planes_alpha_spect_comp}(a), run 
$1P_{0.1}$ is not in equilibrium yet while $1P_{0.2}$ is. For series 3P, according to 
Fig.~\ref{fig:23planes_alpha_spect_comp}(b) run $3P_{0.05}$ is already in 
equilibrium while $3P_{0.01}$ is not.     
A quantitative difference between the two cases could be expected due to the presence
of a larger number of possible 3D triads that can form in series 3P compared to series 1P.  
\\

\begin{figure}[tbp]
\begin{center}
\includegraphics[width=\columnwidth]{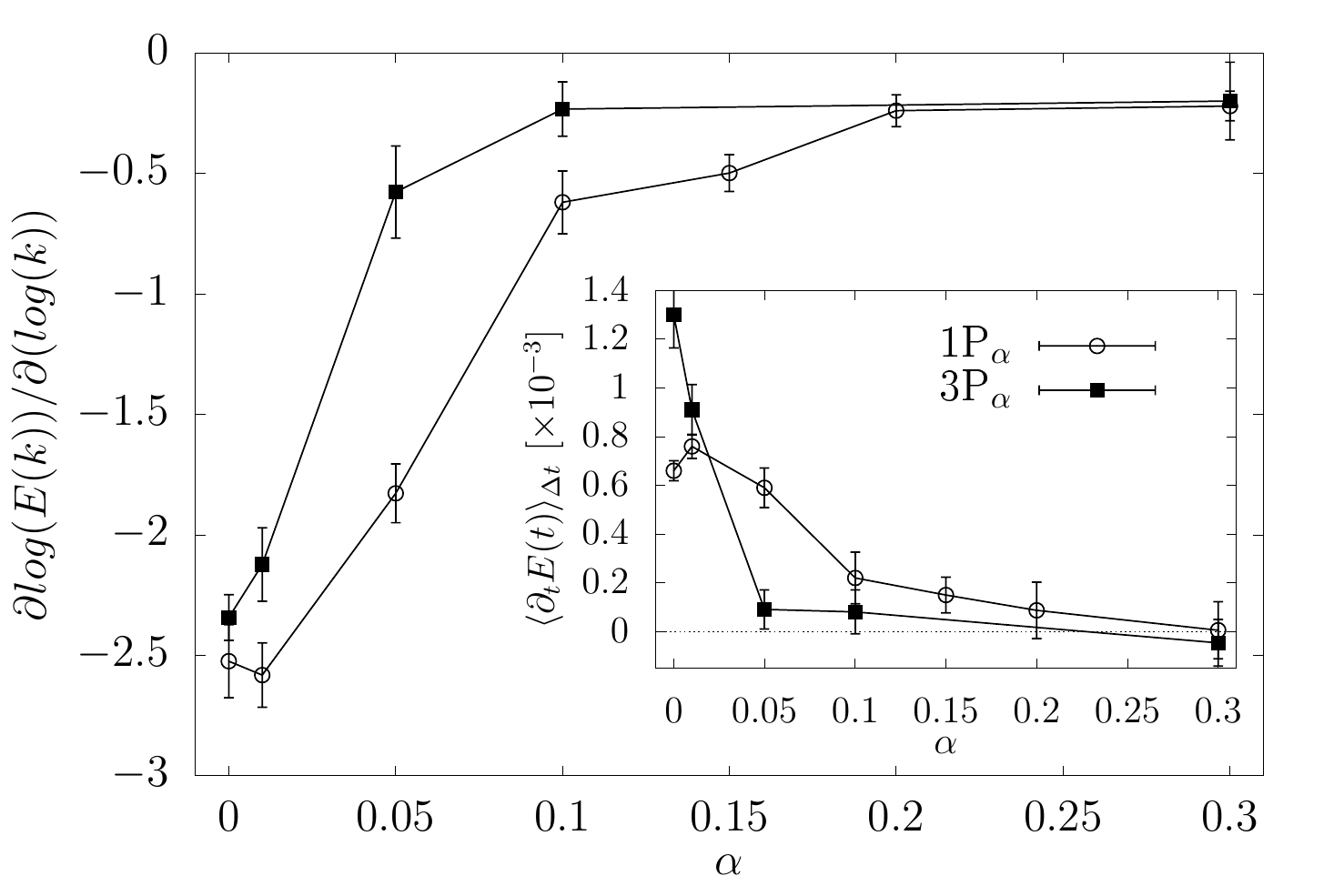} 
\caption{Slopes of $E(k)k^{1-D_{\alpha}(k)}$ shown in Fig.~\ref{fig:23planes_alpha_spect_comp}, in the range $6 \le k \le 16$, as a function of $\alpha$. (Inset) Slope of $E(t)$ shown in Fig.~\ref{fig:23planes_alpha_evol}, 
in the range $13 \leqslant \Delta t_1 \leqslant 27$. The slopes are obtained by least-squares fits and the error bars correspond to the fitting error.
}
\label{fig:23planes_alpha_spectra_slopes}
\end{center}
\end{figure}

\noindent
In order to describe the transition from 2D to 3D turbulence 
more quantitatively, we determine
the deviation of the respective energy spectra 
from absolute equilibrium scaling as a function of $\alpha$. For this purpose 
the slopes of the compensated energy spectra shown in 
Fig.~\ref{fig:23planes_alpha_spect_comp} for different values of $\alpha$ have been 
measured through least-squares fits with results shown in  
Fig.~\ref{fig:23planes_alpha_spectra_slopes} for both series of simulations. 
The results presented in the figure place the value of $\alpha$ at which the transition
occurs in the range $0.01 \leqslant \alpha \leqslant 0.05$ for series 3P and 
in the range $0.1 \leqslant \alpha \leqslant 0.2$ for series 1P.
In Appendix \ref{app:transition_estimate} we present a rough theoretical argument
predicting the critical value $\alpha \simeq 0.13$ based only 
on the geometric structure of the nonlinearities for the $1P_\alpha$ configuration.
The inset of Fig.~\ref{fig:23planes_alpha_spectra_slopes} shows measurements of the time-derivative of the total 
energy $E(t)$ for all runs from series 1P and 3P,
where $\langle \p_t E\rangle =const \neq 0$ 
indicates the presence of an inverse cascade while $\langle \p_t E\rangle \simeq 0$ is satisfied in the 
stationary state. Angled brackets denote a temporal average. Owing to the saturation
effect that occurs eventually even in the base configuration $3P_{0.0}$, the measured values 
of $\langle \p_t E\rangle$ depend on the interval the time-derivative is averaged over. 
\\
\noindent
Visualizations of the flows obtained by adding a fraction $\alpha$ of 3D modes to the two 
base configurations are shown for both series 
in Figs.~\ref{fig:1plane-flux-alpha005}(a)-\ref{fig:3plane-flux-alpha03}(a), where Fig.~\ref{fig:1plane-flux-alpha005}(a) corresponds to a single 2D3C flow with $\alpha =0.05$ ($1P_{0.05}$) while Fig.~\ref{fig:3plane-flux-alpha001}(a) shows to a three coupled 2D3C flows with $\alpha =0.01$ ($3P_{0.01}$).
The respective values of $\alpha$ correspond to perturbed 2D3C flows before the transition to 3D
dynamics has taken place. 
From Fig.~\ref{fig:1plane-flux-alpha005}(a) we observe that the flow now shows
some variations along the $z$-axis, however, the main features of 2D evolution such as 
the formation of large-scale structures, are still visible. In contrast, the coupled 2D3C
flows with $\alpha =0.01$ presented in Fig.~\ref{fig:3plane-flux-alpha001}(a)
shows no such structure, although the characteristic pattern of the coupled 2D3C dynamics
is still visible. 

\subsection{Homo- and heterochiral subfluxes during the 2D-3D transition}
\label{sec:homochiral_numerics}
In this section we investigate the behavior of the energy subfluxes 
corresponding to homo- and heterochiral interactions  during the transition from 
2D to 3D turbulence. According to the arguments presented in 
Sec.~\ref{sec:cascade_directions}, it can be expected that the qualitative behavior of the 
homochiral subfluxes remains unaltered during the transition while the 
heterochiral subflux should visibly change. \\

\begin{figure}[tbp]
\begin{center}
\includegraphics[width=\columnwidth]{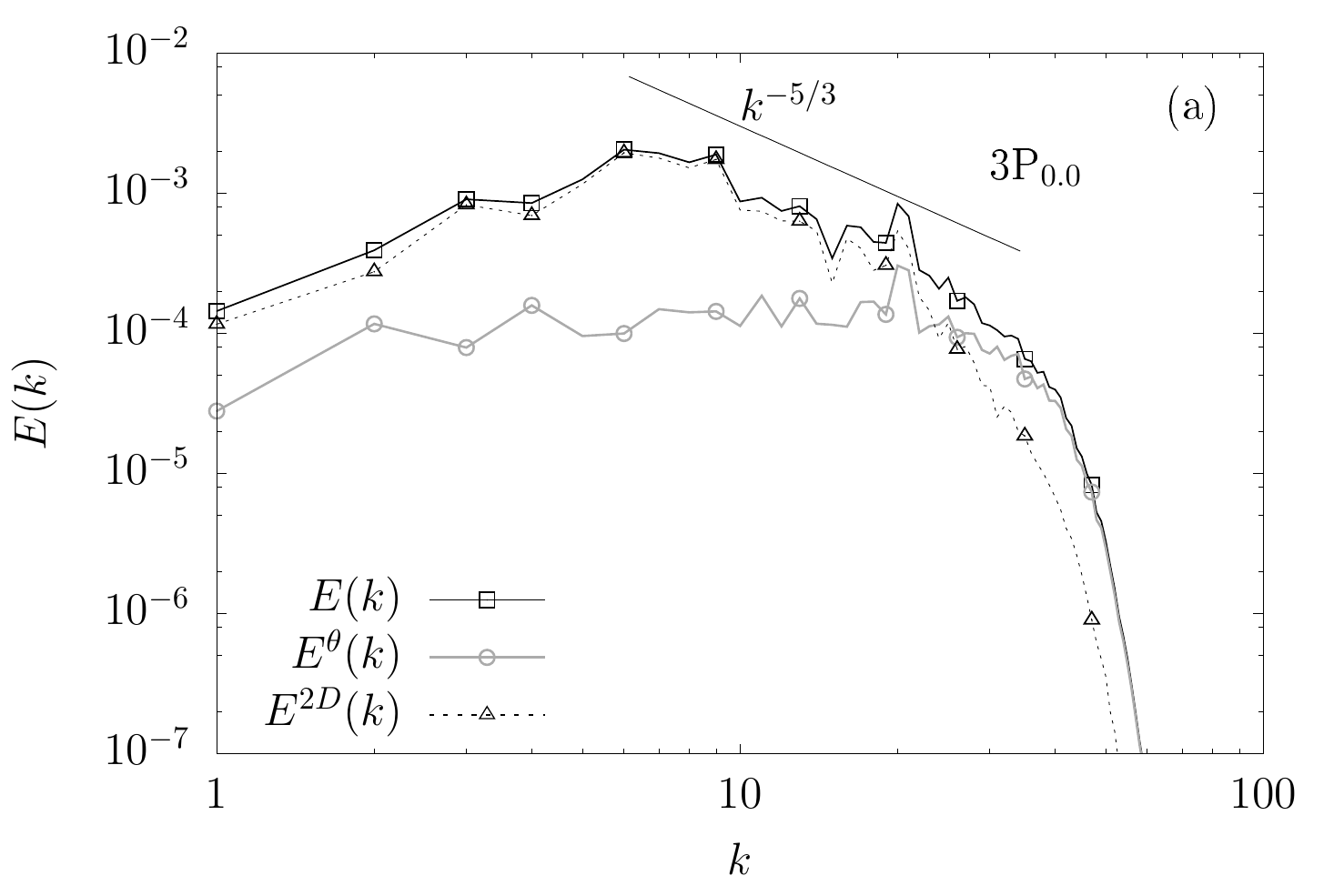} 
\includegraphics[width=\columnwidth]{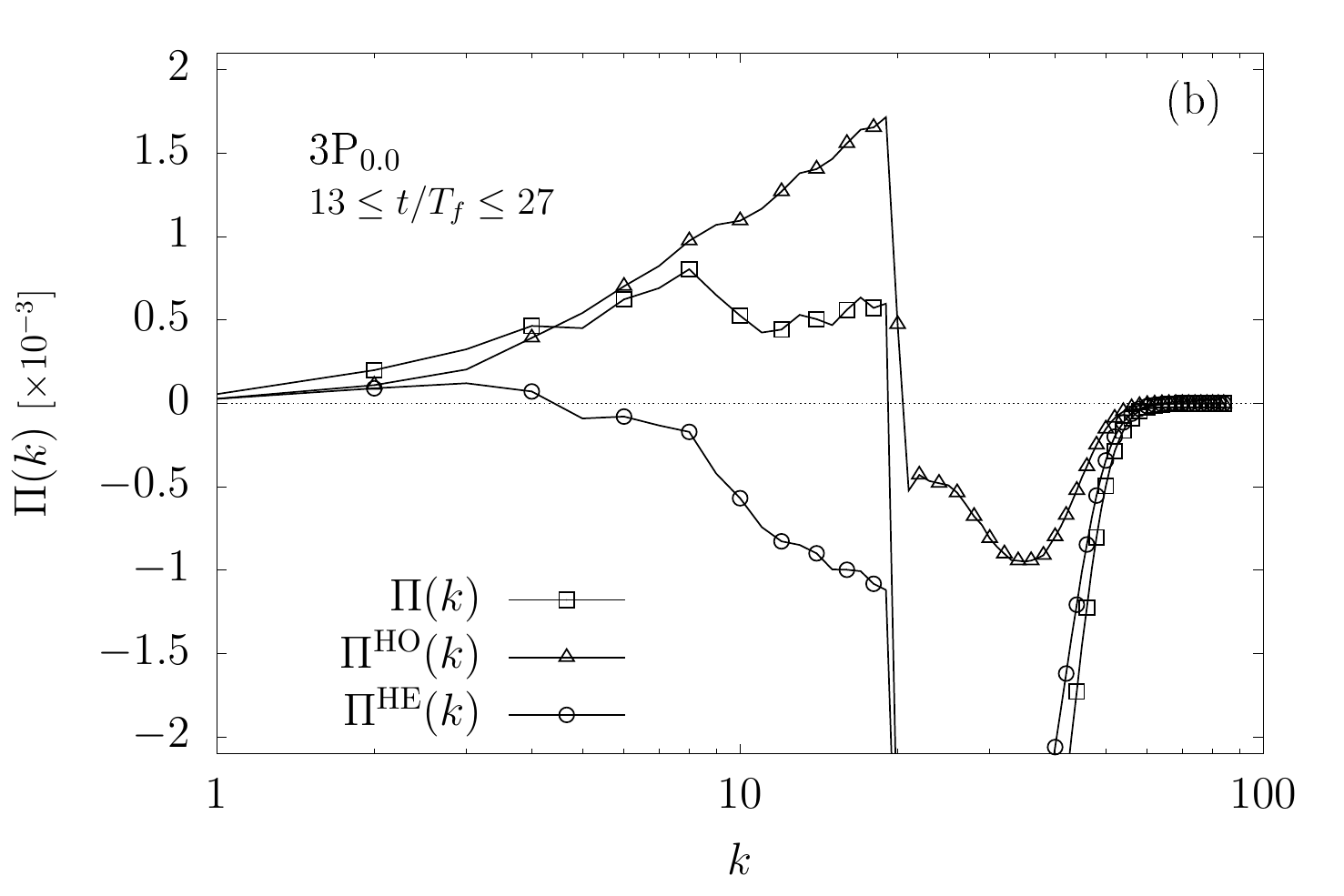} 
\includegraphics[width=\columnwidth]{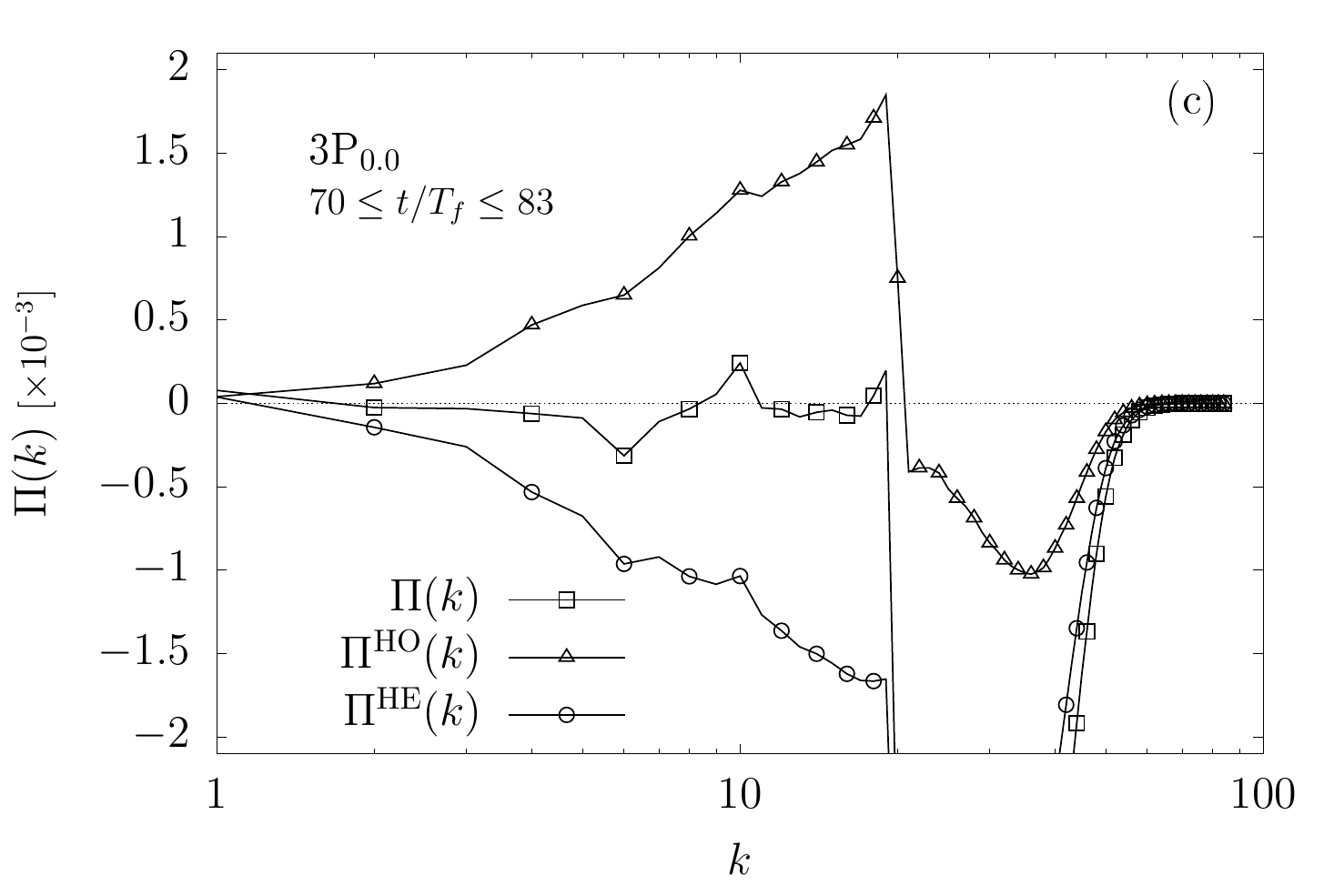} 
\caption{
Three coupled 2D3C flows 3P$_{0.0}$. (a) Energy spectra $E(k)$, $E^{\text{2D}}(k)$ and $E^{\theta}(k)$.
         (b)-(c) Energy flux $\Pi(k)$ and subfluxes $\Pi^{\text{HO}}(k)$, $\Pi^{\text{HE}}(k)$ averaged in the time 
range $t/T_f \in [13,27]$ and $t/T_f \in [70,83]$ respectively.
{
Although $\Pi(k) \simeq 0$ at $k<k_f$ in the statistically steady configuration in panel (c), the
system is out of equilibrium 
since $\Pi(k)$ vanishes owing to competing
forward and backward subfluxes.
}
}
\label{fig:3plane-flux-spectra}
\end{center}
\end{figure}

\noindent
Figure~\ref{fig:3plane-flux-spectra}(a) presents $E^{2D}(k)$ and  
$E^{\theta}(k)$ for one of the three planes. As can be seen, unlike for a single 
2D3C flow shown in Fig.~\ref{fig:1plane-flux-spectra}(a), 
$\theta$ is no longer in equipartition. Furthermore,
$E^{2D}(k)$ now only shows a partial inverse-cascade 
$k^{-5/3}$-scaling at $k<k_f$. Both observations are consistent with the coupled 
2D3C flow becoming partly 3D. 
This is also reflected in the fluxes $\Pi(k)$, $\Pi^{\rm HO}(k)$ and $\Pi^{\rm HE}(k)$ 
shown in 
Fig.~\ref{fig:3plane-flux-spectra}(b) before saturation and in Fig.~\ref{fig:3plane-flux-spectra}(c) 
after saturation. We now observe that $\Pi^{\rm HO}(k)$ dominates the total flux at intermediate 
scales while $\Pi^{\rm HE}(k)$ has changed 
from 2D to 3D behavior, leading to cancellations between the two and hence a lower total inverse 
energy flux compared to the single 2D3C case where all helical combination lead to an inverse cascade 
(see Fig.~\ref{fig:1plane-flux-spectra}(c)). After saturation the homo- and heterochiral subfluxes 
nearly cancel out and the average total inverse flux vanishes.   
Hence the dynamics of the coupled 2D3C flows is consistent with 
a qualitative picture of competing 2D and 3D dynamics, even without a further addition 
of 3D modes in the volume. \\

\begin{figure*}[tpb]
\begin{center}
\includegraphics[width=\columnwidth]{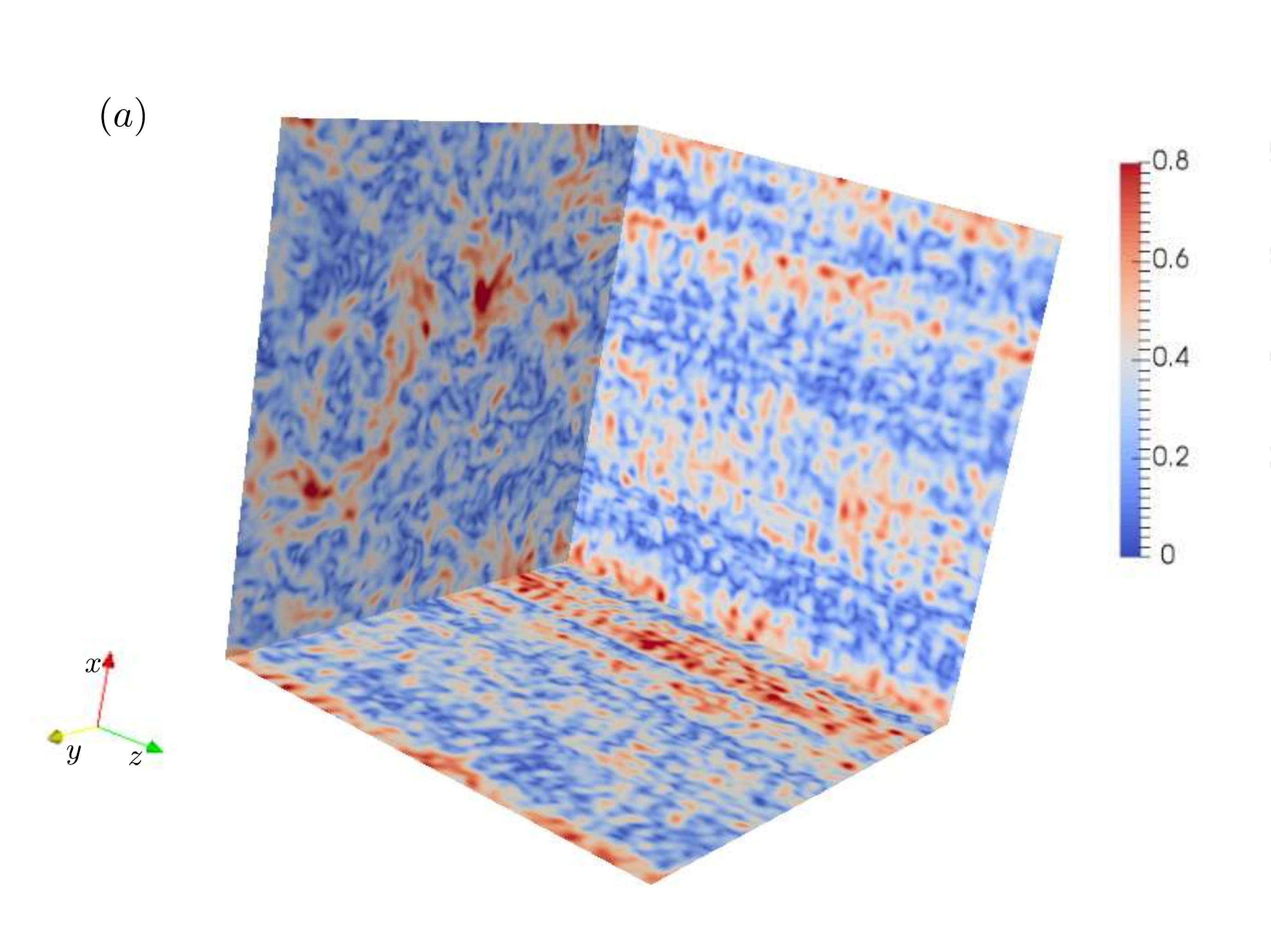} 
\includegraphics[width=\columnwidth]{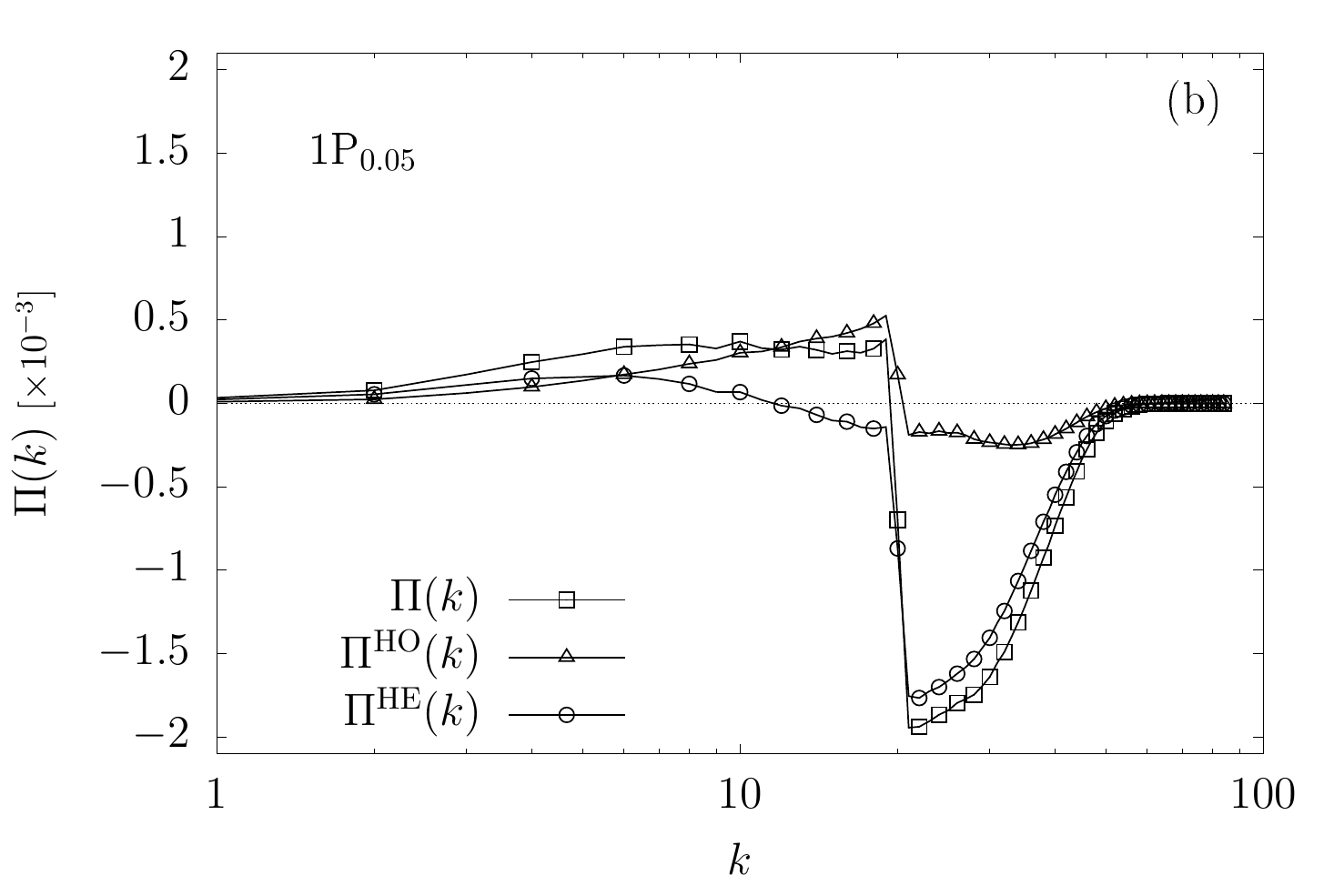}
\caption{
(a) Visualization of \obs{the kinetic energy of} one velocity configuration during the time evolution of 1P$_{0.05}$: 2D3C flow with $\alpha = 0.05$. (b) $\Pi(k)$ and its two different helical components $\Pi^{\text{HO}}(k)$ and $\Pi^{\text{HE}}(k)$ corresponding to 1P$_{0.05}$. 
}
\label{fig:1plane-flux-alpha005}
\end{center}
\end{figure*}

\begin{figure*}[tpb]
\begin{center}
\includegraphics[width=\columnwidth]{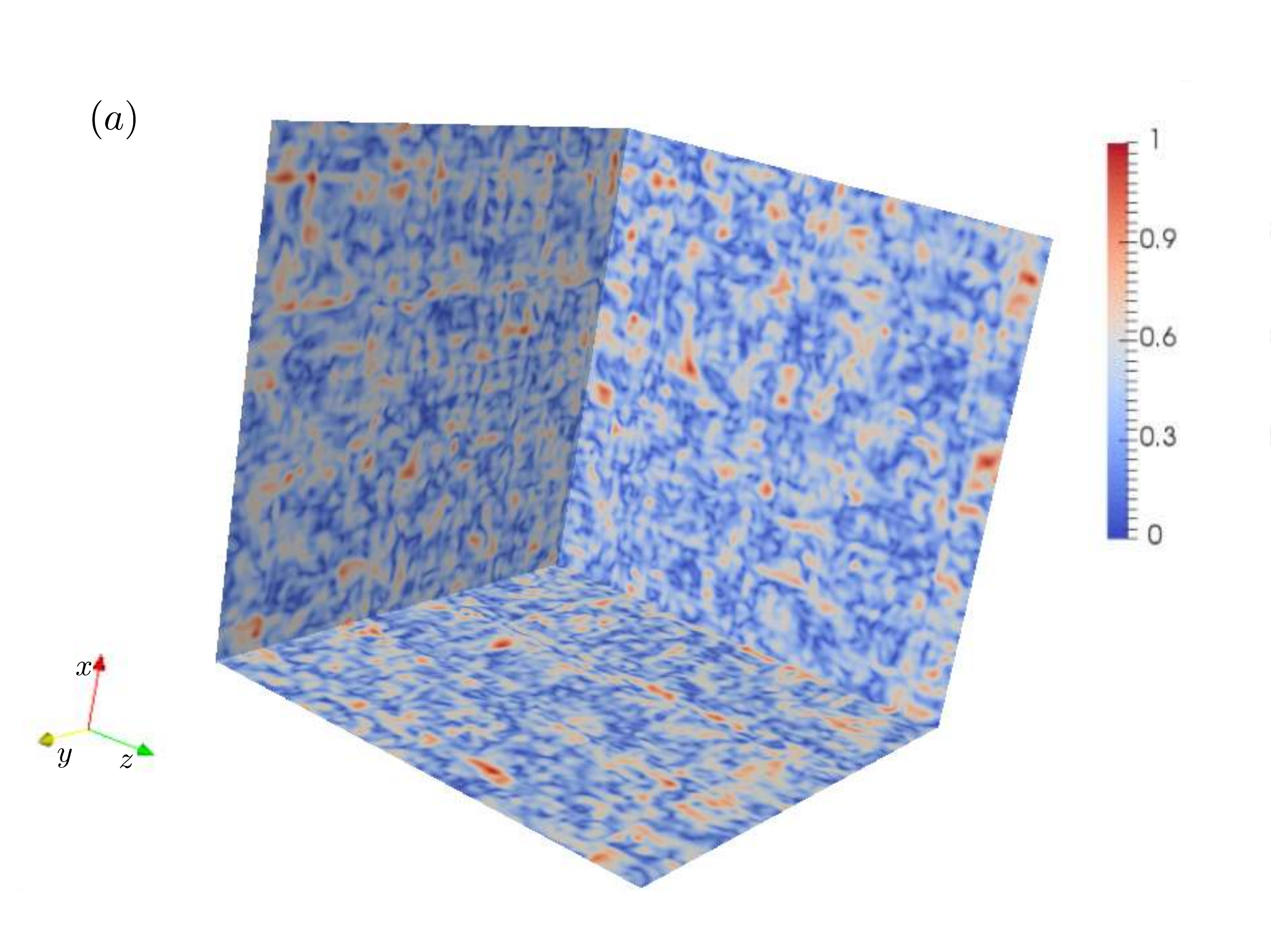} 
\includegraphics[width=\columnwidth]{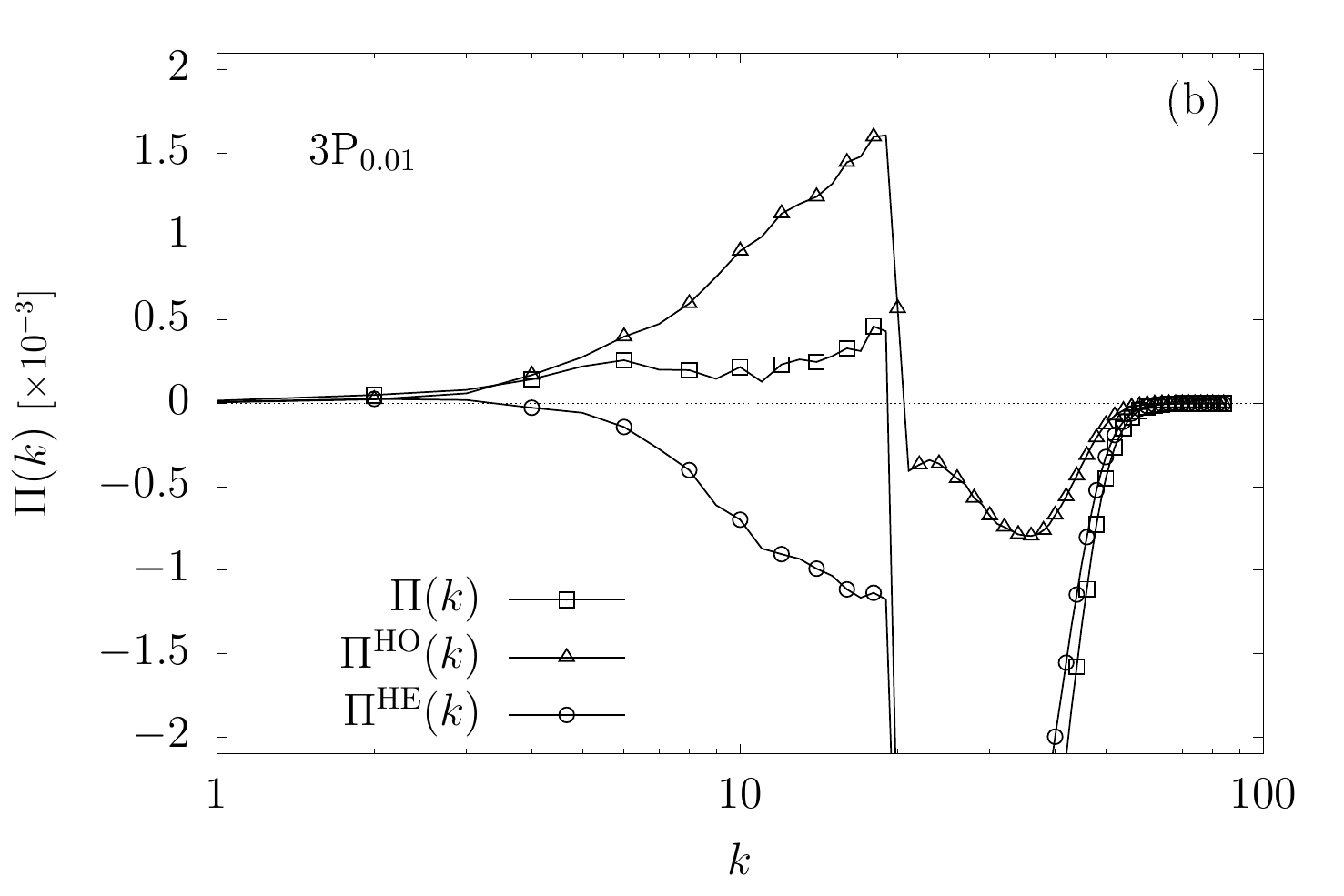} 
\caption{
(a) Visualization of \obs{the kinetic energy of} one velocity configuration during the time evolution of 3P$_{0.01}$: three coupled 2D3C flows with $\alpha = 0.01$. (b) $\Pi(k)$ and its two different helical components $\Pi^{\text{HO}}(k)$ and $\Pi^{\text{HE}}(k)$ corresponding to 3P$_{0.01}$. 
}
\label{fig:3plane-flux-alpha001}
\end{center}
\end{figure*}

\begin{figure*}[tpb]
\begin{center}
\includegraphics[width=\columnwidth]{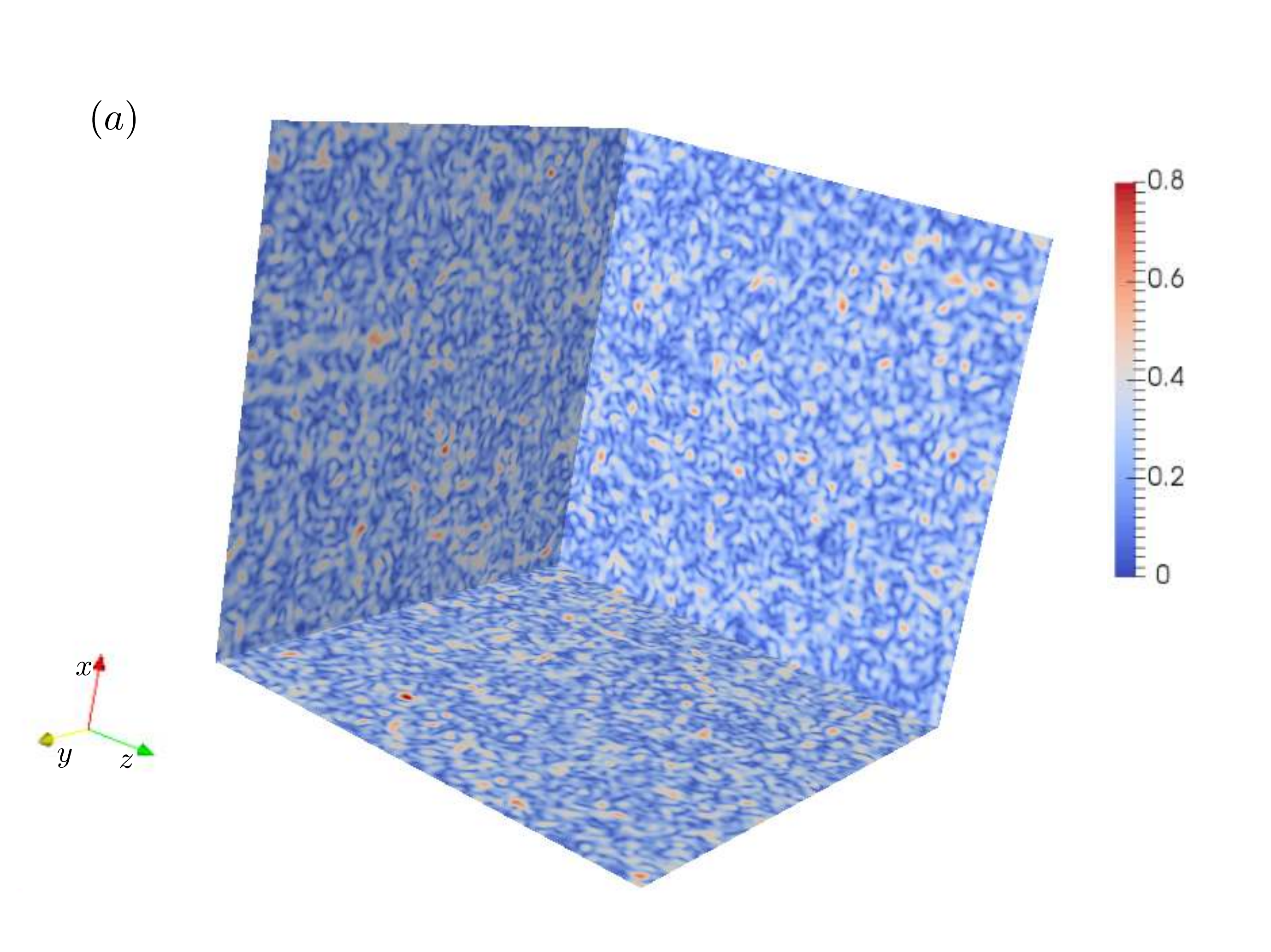} 
\includegraphics[width=\columnwidth]{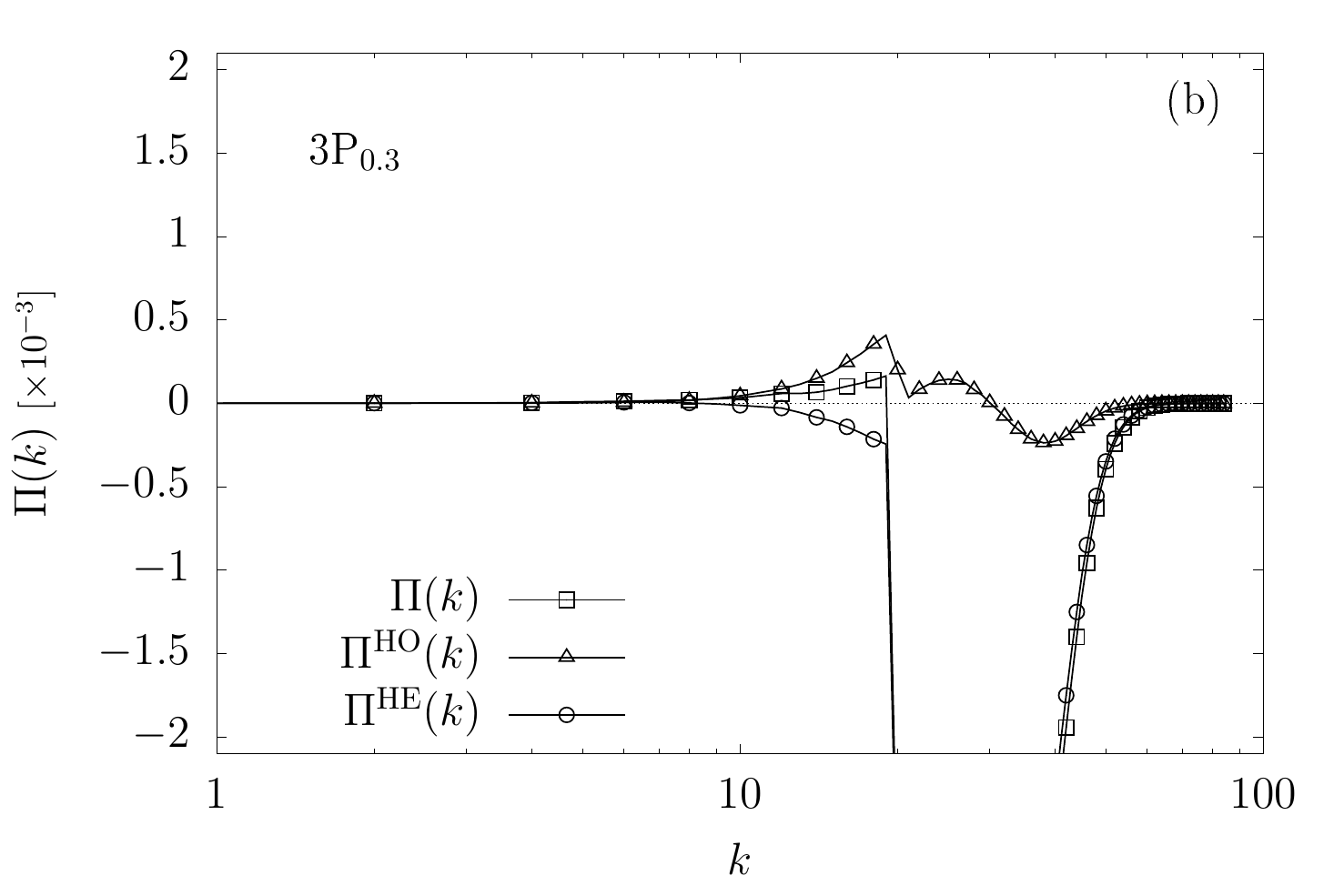}
\caption{
(a) Visualization of \obs{the kinetic energy of} one velocity configuration during the time evolution of 3P$_{0.3}$: three coupled 2D3C flows with $\alpha = 0.3$. (b) $\Pi(k)$ and its two different helical components $\Pi^{\text{HO}}(k)$ and $\Pi^{\text{HE}}(k)$ corresponding to 3P$_{0.3}$.
}
\label{fig:3plane-flux-alpha03}
\end{center}
\end{figure*}

\noindent 
The latter is further supported by the observations made from the total, 
homo- and heterochiral fluxes for runs $1P_{0.05}$ and $3P_{0.01}$ shown in 
Figs.~\ref{fig:1plane-flux-alpha005}(b) and \ref{fig:3plane-flux-alpha001}(b), respectively, where we also find that the 
homochiral flux begins to dominate the total energy flux at intermediate scales larger
than the forcing scale, while the heterochiral flux is changing sign. 
In particular for run $1P_{0.05}$ shown in Fig.~\ref{fig:1plane-flux-alpha005}(b) we 
observe partial 2D and 3D dynamics at different scales. 
In summary, the qualitative behavior of the homochiral dynamics is robust under the transition
from 2D to 3D turbulence, while  the behavior of heterochiral 
energy transfers changes, leading eventually to 
a depletion of the total inverse energy flux. 
{The change of the heterochiral interactions is due
to more and more out-of-plane couplings becoming available 
at the large scales, which according to the results in Sec.~\ref{sec:Fourier} dominate over 
the heterochiral contribution to the inverse energy transfer in the planes.} 
Once the transition to 3D dynamics is complete, all fluxes tend to zero in the wavenumber range $k < k_f$
as shown representatively in Fig.~\ref{fig:3plane-flux-alpha03}(b) for run $3P_{0.3}$. This behavior can 
be expected, since the heterochiral interactions are transferring energy downscale very efficiently, before any upscale energy 
transfer due to homochiral interactions can be established. 
\\

\noindent
{The results on the behavior of homo- and heterochiral interactions 
in isolation suggests that homochiral 2D3C subdynamics enhance the 
2D physics. They are almost 2D in the sense that they conserve the total enstrophy. 
Their dynamical role seems to persists during the 2D3C to 3D transition
despite the system becoming increasingly 3D and conserving 
only the usual 3D inviscid invariants $E$ and $H$.}

\section{Conclusions} \label{sec:conclusions}
We have investigated the structure and the dynamics of 2D3C flows, which can be seen as
the basic building blocks of 3D turbulence, through numerical
simulations and analytical work. Using the Fourier helical decomposition, we
have shown that any 2D3C flow can be described through two stream functions
corresponding to the two helical sectors of the velocity field.  \obs{As} a result, \obs{a} homochiral 2D3C flow is described by one  stream function only. The
projection  onto the submanifold corresponding to the homochiral dynamics enforces a correlation between the component
$\theta \hat{\bm z}$ of the velocity field perpendicular to the $(x,y)$-plane and the
vorticity of the planar component $\omega \hat{\bm z}$. Hence a homochiral 2D3C
flow can never be purely 2D and $\theta$ ceases to obey the same equations of a passive scalar.  The projection operation also
results in the inviscid conservation of the total enstrophy, hence the total 3D kinetic
energy confined to the homochiral submanifold must
display an inverse cascade.  \\
We explore the transition from a  2D3C to a full 3D dynamics by coupling several 2D3C flows
through a set of suitably designed direct numerical 
simulations (DNS).
{
We found that the coupling of three 2D3C flows on mutually orthogonal planes
leads to a stationary regime where $\Pi(k) \simeq 0$ due to competing subfluxes
$\Pi^{HE}(k) \simeq -\Pi^{HO}(k)$. Unlike in full 3D configurations subject 
to small-scale forcing, the kinetic energy is not equally distributed amongst the Fourier modes
at wavenumbers corresponding to scales larger than the forcing scale, and  
we obtain stationary out-of-equilibrium 3D dynamics at the energy containing scales.  
}
The situation is also the opposite of what typically happens in geophysical situations, here the flow is 
2D at high wavenumbers and becomes 3D only at large scales.  
The transition between 2D and 3D turbulence has been further explored through adding a 
percentage of 3D Fourier modes in the whole volume. We found that the homochiral sector tends to transfer energy upward always, while heterochiral
triads start to transfer energy to small scales as soon as a small percentage $\sim 10 \%$ of modes occupy the whole Fourier space isotropically.
We also present a rough argument based on a geometrical balance of 2D-3D triads to predict such a transition. In conclusions, we have shown that it is possible to shed further lights on the entangled dynamics of 2D-2D3C-3D flows by using a decomposition in helical waves and by performing suitable numerical experiments meant to restrict the dynamics on different submanifolds.

\begin{acknowledgments}
\obs{We thank one of the anonymous referees for their interesting remark 
about Ekman friction in the context of the helical decomposition}.
The research leading to these results has received funding from the European
Union's Seventh Framework Programme (FP7/2007-2013) under grant agreement No.
339032. 
We acknowledge very useful discussions with B. Gallet and 
previous collaborations on 
similar topics with R. Benzi, P. Perlekar, G. Sahoo and F. Toschi.
\end{acknowledgments}

\appendix

\section{Geometric factors for 2D and perpendicular evolution equations}
\label{app:coupling_factors}

In order to separate the evolution in the plane from that in the perpendicular direction, we further decompose the
helical basis vectors
\be
\hsk = \hperpk + \hparsk \ ,
\ee
where the projection on the plane
\be
\hperpk \equiv \begin{pmatrix} i{\hat k_y}  \\  -i{\hat k_x} \\ 0 \end{pmatrix} \ ,
\ee
is helicity-independent and $\hat \bk = \bk/k$. The perpendicular component carries
all information on helicity, it is defined as
\be
\hparsk \equiv \begin{pmatrix} 0  \\  0 \\ \obs{s_k} \end{pmatrix} \ ,
\ee
such that $\hparmk = -\hparpk$, and we define $\hpark \equiv \hparpk = \hat \bz$.
The planar and perpendicular components of the evolution equation Eq.~\eqref{eq:Fevolution_psipm} can now be obtained
by taking the inner product of Eq.~\eqref{eq:Fevolution_psipm} with the appropriate basis vector. For the
planar component we obtain
\begin{align}
\label{eq:Fevolution_perp}
\p_t k(\fpsik^+ + \fpsik^-)^* &= 
\frac{1}{2} \sum_{\bk+ \bp+\bq =0} \sum_{s_p,s_q}\frac{p^2-q^2}{k} \nonumber \\
&  \ \ \times  kp \sin\varphi_{k,p}\ \fpsip^{s_p}\fpsiq^{s_q} \ ,
\end{align}
where $\varphi_{k,p}$ is the angle between wavevectors $\bk$ and $\bp$ (see below).
As expected, the coupling factor $\frac{p^2-q^2}{k} kp \sin\varphi_{k,p}$ is helicity-independent. For the perpendicular
component we obtain
\begin{align}
\label{eq:Fevolution_par}
\p_t k(\fpsik^+ - \fpsik^-)^* &= 
\frac{1}{2} \sum_{\bk+ \bp+\bq =0} \sum_{s_p,s_q}(s_pp-s_qq) \nonumber \\
&  \ \ \times kp \sin\varphi_{k,p} \ \fpsip^{s_p}\fpsiq^{s_q} \ ,
\end{align}
where the coupling factor $(s_pp-s_qq) kp \sin\varphi_{k,p}$ now depends on the helicities of the stream functions.\\
The coupling factors are calculated by taking the
inner product of $\hsp \times \hsq$ with either $\hperpk$ (2D) or $\hpark$.
We first calculate
\be
\hsp \times \hsq = 
\begin{pmatrix} i{\hat p_y}  \\  -i{\hat p_x} \\ \obs{s_p} \end{pmatrix}
\times
\begin{pmatrix} i{\hat q_y}  \\  -i{\hat q_x} \\ \obs{s_q} \end{pmatrix} 
= \begin{pmatrix} -is_q{\hat p_x} + is_p {\hat q_x} \\
                -is_q{\hat p_y} + is_p {\hat q_y} \\
                \hat p_y \hat q_x - \hat p_x \hat q_y \end{pmatrix} \ .
\ee
For the 2D geometric factor we obtain
\begin{align}
\hperpk \cdot \left(\hsp \times \hsq \right) &=
\begin{pmatrix} i{\hat k_y}  \\  -i{\hat k_x} \\ 0 \end{pmatrix} 
\cdot
\begin{pmatrix} -is_q{\hat p_x} + is_p {\hat q_x} \\ 
                -is_q{\hat p_y} + is_p {\hat q_y} \\ 
                \hat p_y \hat q_x - \hat p_x \hat q_y \end{pmatrix} \nonumber \\
&= -\left( \hat \bk \times (s_q \hat \bp -s_p \hat \bq)\right)_z \ ,
\end{align}
such that the coupling factor in front of $\fpsip^{s_p} \fpsiq^{sq}$ in
Eq.~\eqref{eq:Fevolution_perp} becomes
\begin{align}
(s_pp-s_qq) \  pq \ \hperpk \cdot & \left(\hsp \times \hsq \right) \nonumber \\ 
&=-(s_pp-s_qq) \left( \hat \bk \times (s_qq \bp -s_pp \bq)\right)_z \nonumber \\ 
&=-(s_pp-s_qq) \left( \hat \bk \times (s_qq +s_pp) \bp\right)_z \nonumber \\ 
&=-\frac{p^2-q^2}{k} kp \sin\varphi_{k,p} \ , 
\end{align}
where the triad condition $\bk+\bp+\bq =0$ was used in the second step.
For the geometric factor in the perpendicular component we obtain
\be
\hpark \cdot \left(\hsp \times \hsq \right) =
\begin{pmatrix} 0  \\  0 \\ 1 \end{pmatrix}
\cdot
\begin{pmatrix} -is_q{\hat p_x} + is_p {\hat q_x} \\
                -is_q{\hat p_y} + is_p {\hat q_y} \\
                \hat p_y \hat q_x - \hat p_x \hat q_y \end{pmatrix}
= -\left( \hat \bp \times \hat \bq \right)_z \ ,
\ee
such that the coupling factor in front of $\fpsip^{s_p} \fpsiq^{sq}$ in
Eq.~\eqref{eq:Fevolution_par} becomes
\begin{align}
(s_pp-s_qq) \ & pq \ \hpark \cdot \left(\hsp \times \hsq \right) 
=-(s_pp-s_qq) (\bp \times \bq)_z \nonumber \\
&=-(s_pp-s_qq) (\bk \times \bp)_z \nonumber \\
&=-(s_pp-s_qq)kp \sin\varphi_{k,p} \ , 
\end{align}
where the triad condition $\bk+\bp+\bq =0$ was used in the second step.

\section{Passive scalar evolution in 2D turbulence} \label{app:passive_scalar}
The helical decomposition of the 2D3C flow can formally be applied to the dynamics of a passive scalar $\theta$ in 2D
turbulence by defining
\be
\psi^+ \equiv \frac{1}{2}(\psi + (-\Delta)^{-1/2}\theta)
\quad \mbox{and} \quad
\psi^- \equiv \frac{1}{2}(\psi - (-\Delta)^{-1/2}\theta) \ ,
\ee
where $\psi$ is the stream function of the 2D flow. The decomposition in Fourier space is then
\be
\fpsik^+ = \frac{1}{2}(\fpsik + \fthetak/k)
\quad \mbox{and} \quad
\fpsik^- = \frac{1}{2}(\fpsik - \fthetak/k) \ .
\ee
The Fourier-space evolutions equations for $\fpsik$ and $\fthetak$
\begin{align}
\label{eq:Fevolution_2d}
\p_t \fpsik^* &=\sum_{\bk+ \bp+\bq =0} \frac{p^2-q^2}{k^2} kp \sin\varphi_{k,p} \ \fpsip\fpsiq \ , \\ 
\label{eq:Fevolution_theta}
\p_t \fthetak^* &=\frac{1}{2} \sum_{\bk+ \bp+\bq =0} kp \sin\varphi_{k,p}(\fpsip\fthetaq-\fpsiq\fthetap) \ ,
\end{align}
correspond in the helical decomposition to Eqs.~\eqref{eq:Fevolution_perp}
and \eqref{eq:Fevolution_par}, respectively. Following the discussion in Sec.~\ref{sec:cascade_directions},
the passive scalar evolution is therefore dominated by
`heterochiral' interactions, which according to the stability arguments in Ref.~\onlinecite{Waleffe92} would
indicate that $E^\theta$ should display a forward cascade.
We note that a direct stability analysis of single-triad dynamical systems derived from
Eqs.~\eqref{eq:Fevolution_2d} and \eqref{eq:Fevolution_theta} in conjunction is not possible
because the usual trick using $2^{nd}$-order time derivatives does not lead to closed equations for the
passive scalar. However, for an active scalar it may work.
This problem does not arise for triads involving only the stream function, 
the stability analysis for the 2D dynamics 
has been carried out in Ref.~\onlinecite{Waleffe92}.

\section{2D-3D transition: Attempt to calculate percentage of 3D triads necessary for transition}
\label{app:transition_estimate}
Assuming that the direction of the cascade is determined by the geometrical constraints {\em only}, 
it is possible to estimate the percentage of 3D triads that need to be active in order to
change from 2D to 3D turbulence. This assumption is quite drastic, hence the results can only 
serve as guidance. 
\\  

\noindent
The aim is to calculate the
flux coming from the 2D dynamics and the fraction of added 
3D-triads and determine when it changes sign.
Following the ideas introduced by Kraichnan \cite{Kraichnan67} and Waleffe \cite{Waleffe92}, 
we consider the energy flux in the inertial range 
(see Refs.~\onlinecite{Kraichnan67,Waleffe92} for details)
\begin{align}
\label{eq:flux-2d-3d}
&\Pi(k) = \sum_{i=1}^8\int_0^1 dv \int_1^{1+v} \hspace{-1em}dw \nonumber \\
&\times \Big[
(1-\alpha)\underbrace{\left (\frac{(v^2-1)\ln{w} + (1-w^2)\ln{v}}{w^2-v^2} \right) T^{(i)}_{2D}(1,v,w)}_\text{plane} \nonumber \\
&+ \alpha\underbrace{\left(\frac{(s_vv-s_11)\ln{w}+(s_11-s_ww)\ln{v}}{s_ww-s_vv}\right)T^{(i)}_{3D}(1,v,w)}_\text{added 3D triads}\Big] \, 
\end{align}
where the superscript $(i)$ labels the eight helical interactions and
where we have introduced an adjustable parameter $0\leqslant \alpha \leqslant 1$, such that $\alpha =
1$ corresponds to the full 3D Navier-Stokes equation and $\alpha = 0$ to 2D
evolution.  The factor $(1-\alpha)$ in front of the 2D evolution term is necessary in order to
avoid double-counting the plane.  
Equation ~\eqref{eq:flux-2d-3d} reduces to 
the exact expression of the inertial-range energy
flux in 3D isotropic turbulence for $\alpha =1$ and to that for the energy flux
in 2D isotropic turbulence for $\alpha = 0$. The reason for being able to 
formally superpose the two expressions for fractional values of $\alpha$  
is that the inertial-range scaling exponents in 2D and 3D are the 
same, and that the differences in the vectorial character of the coupling
between 2D and 3D have been absorbed into the respective 
factors in front of the terms $T^{(i)}_{2D}(1,v,w)$ and 
$T^{(i)}_{3D}(1,v,w)$.    
The term $T^{(i)}_{2D}(1,v,w)$ in Eq.~\eqref{eq:flux-2d-3d} in fact does not
depend on helicity, and according to arguments based on statistical mechanics \cite{Kraichnan67} or 
on the stability of equilibria of single-triad 2D dynamical systems \cite{Waleffe92} $T^{(i)}_{2D}(1,v,w) \leqslant 0$ for all $i$. The sign of the 3D evolution
term $T^{(i)}_{3D}(1,v,w)$ changes depending on the type of helical interaction \cite{Waleffe92}. 
For homochiral interactions we have
$T^{(i)}_{3D}(1,v,w)<0$ as in 2D, and heterochiral interactions with 
$s_1 \neq s_w$ lead to
$T^{(i)}_{3D}(1,v,w)>0$ while those with $s_v \neq s_1 = s_w$ 
lead to $T^{(i)}_{3D}(1,v,w)<0$. 
In order to be able to calculate the integral, 
we now assume that $|T^{(i)}(1,v,w)| = |T^{(j)}(1,v,w)|$ for $j \neq i$, 
as no explicit expression for $T^{(i)}(1,v,w)$ is available. 
In addition, we further assume
that the magnitude of the transfer term is independent of the geometry
of the triad, hence in the following we set
$\mathcal{T} \equiv |T^{(i)}(1,v,w)|$ for all interactions $(i)$ and all
$v \leqslant 1 \leqslant w \leqslant 1+v$, and we absorb 
that the aforementioned sign changes 
into the sum over the geometric factors. 
The integrand in Eq.~\eqref{eq:flux-2d-3d} can then be simplified and within the approximations made,
the equation can be written as
\begin{align}
\Pi(k) &\simeq 2\mathcal{T}\int_0^1 dv \int_1^{1+v}\hspace{-1em}dw \frac{1}{w^2-v^2}\nonumber \\
&\times \big( 4(1-\alpha)[(v^2-1)\ln{w}+(1-w^2)\ln{v}] \nonumber \\
& \ \ +\alpha[4w(\ln{v}-\ln{w})-w(w-v)\ln{v}]\big) \ ,
\end{align}
where we used mirror symmetry to remove the sum over all helical interactions.
Now $\Pi(k)=0$ as a function of $\alpha$ gives a rough 
estimate of the value of $\alpha$ necessary for a change
in the sign of the flux and therefore in the cascade direction.
Evaluation of the integral yields $\alpha \simeq 0.1275$ for $\Pi(k)\simeq 0$.
This value corresponds to a transition only due to the geometry of the
nonlinear coupling, as we assumed that all helical couplings 
are of the same magnitude which in reality may not be the case. 

\bibliographystyle{unsrt}
\bibliography{refs,refs_les}

\begin{thebibliography}{10}

\bibitem{Kraichnan67}
R.~H. Kraichnan.
\newblock {Inertial ranges in two-dimensional turbulence}.
\newblock {\em Phys. Fluids}, 10(7):1417, 1967.

\bibitem{Lilly69}
D.~K. Lilly.
\newblock {Numerical simulation of two-dimensional turbulence}.
\newblock {\em Phys. Fluids}, 12:{II--240}--{II--249}, 1969.

\bibitem{Sommeria86}
J.~Sommeria.
\newblock {Experimental study of the two-dimensional inverse energy cascade in
  a square box}.
\newblock {\em J. Fluid Mech.}, 170:139--168, 1986.

\bibitem{Smith93}
L.~Smith and V.~Yakhot.
\newblock {Bose condensation and small-scale structure generation in a random
  force driven 2D turbulence}.
\newblock {\em Phys. Rev. Lett.}, 71:352--355, 1993.

\bibitem{Paret97}
J.~Paret and P.~Tabeling.
\newblock {Experimental observation of the two-dimensional inverse energy
  cascade}.
\newblock {\em Phys. Rev. Lett.}, 79:4162--4165, 1997.

\bibitem{Gotoh98}
T.~Gotoh.
\newblock {Energy spectrum in the inertial and dissipation ranges of
  two-dimensional steady turbulence}.
\newblock {\em Phys. Rev. E}, 57:2984--2991, 1998.

\bibitem{Chen03}
S.~Chen, R.~Ecke, G.~Eyink, X.~Wang, and Z.~Xiao.
\newblock {Physical mechanism of the two-dimensional enstrophy cascade}.
\newblock {\em Phys. Rev. Lett.}, 91:214501, 2003.

\bibitem{Lindborg00}
E.~Lindborg and K.~Alvelius.
\newblock {The kinetic energy spectrum of the two-dimensional enstrophy
  turbulence cascade}.
\newblock {\em Phys. Fluids}, 12:945--947, 2000.

\bibitem{Hossain83}
M.~Hossain, W.~Matthaeus, and D.~Montgomery.
\newblock {Long-time states of inverse cascades in the presence of a maximum
  length scale}.
\newblock {\em J. Plasma Phys.}, 30:479--493, 1983.

\bibitem{Smith94}
L.~Smith and V.~Yakhot.
\newblock {Finite-size effects in forced two-dimensional turbulence}.
\newblock {\em J. Fluid Mech.}, 274:115--138, 1994.

\bibitem{Paret98}
J.~Paret and P.~Tabeling.
\newblock {Intermittency in the two-dimensional inverse cascade of energy:
  experimental observations}.
\newblock {\em Phys. Fluids}, 10:3126--3136, 1998.

\bibitem{Boffetta10}
G.~Boffetta and S.~Musacchio.
\newblock {Evidence for the double cascade scenario in two-dimensional
  turbulence}.
\newblock {\em Phys. Rev. E}, 82:016307, 2010.

\bibitem{Boffetta12}
G.~Boffetta and R.~E. Ecke.
\newblock {Two-dimensional turbulence}.
\newblock {\em Annu. Rev. Fluid Mech.}, 44:427--451, 2012.

\bibitem{Frisch95}
U.~Frisch.
\newblock {\em Turbulence: the legacy of A. N. Kolmogorov}.
\newblock Cambridge University Press, 1995.

\bibitem{Celani10}
Antonio Celani, Stefano Musacchio, and Dario Vincenzi.
\newblock Turbulence in more than two and less than three dimensions.
\newblock {\em Phys. Rev. Lett.}, 104:184506, 2010.

\bibitem{Xia11}
H.~Xia~D. Byrne, G.~Falkovich, and M.~Shats.
\newblock {Upscale energy transfer in thick turbulent fluid layers}.
\newblock {\em Nat. Phys.}, 7:321--324, 2011.

\bibitem{Nastrom84}
G.~D. Nastrom, K.~S. Gage, and W.~H. Jasperson.
\newblock Kinetic energy spectrum of large- and mesoscale atmospheric
  processes.
\newblock {\em Nature}, 310:36--38, 1984.

\bibitem{Benavides17}
S.~J. Benavides and A.~Alexakis.
\newblock {Critical Transitions in Thin Layer Turbulence}.
\newblock {\em J. Fluid Mech.}, 822:364--385, 2017.

\bibitem{Cambon89}
C.~Cambon and L.~Jacquin.
\newblock {Spectral approach to non-isotropic turbulence subjected to
  rotation}.
\newblock {\em J. Fluid Mech.}, 202:295--317, 1989.

\bibitem{Waleffe93}
F.~Waleffe.
\newblock Inertial transfers in the helical decomposition.
\newblock {\em Phys. Fluids A}, 5:677--685, 1993.

\bibitem{Smith99}
L.~M. Smith and F.~Waleffe.
\newblock {Transfer of energy to two-dimensional large scales in forced,
  rotating three-dimensional turbulence}.
\newblock {\em Phys. Fluids}, 11:1608, 1999.

\bibitem{Chen05}
Q.~Chen, S.~Chen, G.~L. Eyink, and D.~D. Holm.
\newblock {Resonant interactions in rotating homogeneous three-dimensional
  turbulence}.
\newblock {\em J. Fluid Mech.}, 542:139--164, 2005.

\bibitem{Mininni09}
Pablo~D. Mininni, Alexandros Alexakis, and Annick Pouquet.
\newblock Scale interactions and scaling laws in rotating flows at moderate
  rossby numbers and large reynolds numbers.
\newblock {\em Phys. Fluids}, 21:015108, 2009.

\bibitem{Gallet15a}
B.~Gallet.
\newblock {Exact two-dimensionalization of rapidly rotating
  large-Reynolds-number flows}.
\newblock {\em J. Fluid Mech.}, 783:412--447, 2015.

\bibitem{BiferalePRX2016}
L.~Biferale, F.~Bonaccorso, I.~M. Mazzitelli, M.~A.~T. {van Hinsberg}, A.~S.
  Lanotte, S.~Musacchio, P.~Perlekar, and F.~Toschi.
\newblock {Coherent Structures and Extreme Events in Rotating Multiphase
  Turbulent Flows}.
\newblock {\em Phys. Rev. X}, 6:041036, 2016.

\bibitem{Moffatt67}
H.~K. Moffatt.
\newblock On the suppression of turbulence by a uniform magnetic field.
\newblock {\em J. Fluid Mech.}, 28:571--592, 1967.

\bibitem{Alemany79}
A.~Alemany, R.~Moreau, P.~L. Sulem, and U.~Frisch.
\newblock {Influence of an external magnetic field on homogeneous MHD
  turbulence}.
\newblock {\em J. M\'ec.}, 18:277--312, 1979.

\bibitem{Zikanov98}
O.~Zikanov and A.~Thess.
\newblock Direct numerical simulation of forced mhd turbulence at low magnetic
  reynolds number.
\newblock {\em J. Fluid Mech.}, 358:299¿333, 1998.

\bibitem{Gallet15}
B.~Gallet and C.~R. Doering.
\newblock {Exact two-dimensionalization of low-magnetic-Reynolds-number flows
  subject to a strong magnetic field}.
\newblock {\em J. Fluid Mech.}, 773:154--177, 2015.

\bibitem{Alexakis11}
Alexandros Alexakis.
\newblock Two-dimensional behavior of three-dimensional magnetohydrodynamic
  flow with a strong guiding field.
\newblock {\em Phys. Rev. E}, 84:056330, 2011.

\bibitem{Bigot11}
Barbara Bigot and S\'ebastien Galtier.
\newblock Two-dimensional state in driven magnetohydrodynamic turbulence.
\newblock {\em Phys. Rev. E}, 83:026405, 2011.

\bibitem{Waleffe92}
F.~Waleffe.
\newblock The nature of triad interactions in homogeneous turbulence.
\newblock {\em Phys. Fluids A}, 4:350--363, 1992.

\bibitem{Biferale12}
L.~Biferale, S.~Musacchio, and F.~Toschi.
\newblock {Inverse energy cascade in three-dimensional isotropic turbulence}.
\newblock {\em Phys. Rev. Lett.}, 108:164501, 2012.

\bibitem{Alexakis17}
Alexandros Alexakis.
\newblock Helically decomposed turbulence.
\newblock {\em Journal of Fluid Mechanics}, 812:752--770, 2017.

\bibitem{Moffatt14}
H.~K. Moffatt.
\newblock Note on the triad interactions of homogeneous turbulence.
\newblock {\em J. Fluid Mech.}, 741:R3, 2014.

\bibitem{Constantin88}
P.~Constantin and A.~Majda.
\newblock {The Beltrami spectrum for incompressible flows}.
\newblock {\em Commun. Math. Phys.}, 115:435--456, 1988.

\bibitem{Biferale13}
L.~Biferale and E.~S. Titi.
\newblock {On the global regularity of a helical-decimated version of the 3D
  Navier-Stokes equation}.
\newblock {\em Journ. Stat. Phys}, 151:1089, 2013.

\bibitem{Sahoo15}
G.~Sahoo, F.~Bonaccorso, and L.~Biferale.
\newblock {Role of helicity for large- and small-scale turbulent fluctuations}.
\newblock {\em Phys. Rev. E}, 92:051002, 2015.

\bibitem{DePietro15}
M.~{De Pietro}, L.~Biferale, and A.~A. Mailybaev.
\newblock {Inverse energy cascade in nonlocal helical shellmodels of
  turbulence}.
\newblock {\em Phys. Rev. E}, 92:043021, 2015.

\bibitem{Rathmann16}
N.~M. Rathmann and P.~D. Ditlevsen.
\newblock Pseudo-invariants contributing to inverse energy cascades in
  three-dimensional turbulence.
\newblock {\em Phys. Rev. Fluids}, 2:054607, 2017.

\bibitem{Falkovich01}
G.~Falkovich, K.~Gaw\ifmmode~\mbox{\c{e}}\else \c{e}\fi{}dzki, and
  M.~Vergassola.
\newblock Particles and fields in fluid turbulence.
\newblock {\em Rev. Mod. Phys.}, 73:913--975, 2001.

\bibitem{Patterson71}
G.~S. Patterson and S.~A. Orszag.
\newblock {Spectral {C}alculations of {I}sotropic {T}urbulence: {E}fficient
  {R}emoval of {A}liasing {I}nteractions}.
\newblock {\em Phys. Fluids}, 14:2538--2541, 1971.

\bibitem{Frisch12}
Uriel Frisch, Anna Pomyalov, Itamar Procaccia, and Samriddhi~Sankar Ray.
\newblock {Turbulence in Noninteger Dimensions by Fractal Fourier Decimation}.
\newblock {\em Phys. Rev. Lett.}, 108:074501, 2012.

\bibitem{Lanotte15}
Alessandra~S. Lanotte, Roberto Benzi, Shiva~K. Malapaka, Federico Toschi, and
  Luca Biferale.
\newblock {Turbulence on a Fractal Fourier Set}.
\newblock {\em Phys. Rev. Lett.}, 115:264502, 2015.

\bibitem{Buzzicotti16}
Michele Buzzicotti, Akshay Bhatnagar, Luca Biferale, Alessandra~S. Lanotte, and
  Samriddhi~Sankar Ray.
\newblock {Lagrangian statistics for Navier–Stokes turbulence under
  Fourier-mode reduction: fractal and homogeneous decimations}.
\newblock {\em New J. Physics}, 18:113047, 2016.

\bibitem{Dallasetal15}
V.~Dallas, S.~Fauve, and A.~Alexakis.
\newblock {Statistical Equilibria of Large Scales in Dissipative Hydrodynamic
  Turbulence}.
\newblock {\em Phys. Rev. Lett}, 115:204501, 2015.

\bibitem{Hopf52}
E.~Hopf.
\newblock {Statistical {Hydromechanics} and {F}unctional {C}alculus}.
\newblock {\em Indiana Univ. Math. J.}, 1:87 -- 123, 1952.

\bibitem{Kraichnan73}
Robert~H. Kraichnan.
\newblock Helical turbulence and absolute equilibrium.
\newblock {\em J. Fluid Mech.}, 59:745--752, 1973.

\end{thebibliography}

\end{document}